\documentclass[acmsmall]{acmart}

\usepackage{fancyhdr}
\usepackage[normalem]{ulem}
\usepackage{amsmath}
\usepackage{graphics}
\usepackage{graphicx}
\usepackage{multirow} 
\usepackage{epsfig}
\usepackage{epstopdf}
\usepackage{url}
\usepackage{color}
\usepackage{siunitx}
\usepackage{algorithm}
\usepackage{algpseudocode}
\usepackage[us,12hr]{datetime}
\usepackage{subcaption}
\usepackage{xspace}
\usepackage{setspace}
\usepackage{pifont}
\usepackage{enumitem}
\usepackage{refcount}
\usepackage{calc}

\newcommand{\squishlist} {
    \begin{list}{$\bullet$} {
        \setlength{\itemsep}{-2pt}
        \setlength{\parsep}{2pt}
        \setlength{\topsep}{0pt}
        \setlength{\partopsep}{0pt}
        \setlength{\leftmargin}{1.0em}
        \setlength{\labelwidth}{1em}
        \setlength{\labelsep}{0.5em}
    }
}

\newcommand{\squishend} {
    \end{list}
}

\usepackage{soul}
\soulregister\cite7
\soulregister\ref7
\soulregister\pageref7

\usepackage{tikz}
\newcommand*\circled[1]{\tikz[baseline=(char.base)]{
            \node[shape=circle,draw,inner sep=.4pt,fill=black, text=white] (char) {#1};}}

\fancypagestyle{firstpage}{
  \fancyhf{}

 \pagenumbering{arabic}
}

\setcopyright{none}
\renewcommand\footnotetextcopyrightpermission[1]{}
\makeatletter
\patchcmd{\@maketitle}
  {\addvspace{0.5\baselineskip}\egroup}
  {\addvspace{-2\baselineskip}\egroup}
  {}
  {}
\makeatother

\begin{document}
%%%%%%%%%%%---SETME-----%%%%%%%%%%%%%
\title[Refresh Triggered Computation]
{Refresh Triggered Computation: Improving the Energy Efficiency \\ of
Convolutional Neural Network Accelerators} 
%%%%%%%%%%%%%%%%%%%%%%%%%%%%%%%%%%%%

\author{Syed M. A. H. Jafri}
\affiliation{%
    \institution{KTH Royal Institute of Technology}
}

\author{Hasan Hassan}
\affiliation{%
    \institution{ETH Z{\"u}rich}
}

\author{Ahmed Hemani}
\affiliation{%
    \institution{KTH Royal Institute of Technology}
}

\author{Onur Mutlu}
\affiliation{%
    \institution{ETH Z{\"u}rich}
}

\renewcommand{\shortauthors}{Jafri, et al.}

%%%%%% -- PAPER CONTENT STARTS-- %%%%%%%%
%\setstretch{0.985}

\begin{abstract}

    To employ a Convolutional Neural Network (CNN) in an energy-constrained
    embedded system, it is critical for the CNN implementation to be highly
    energy efficient. Many recent studies propose CNN accelerator
    architectures with custom computation units that try to improve
    energy-efficiency and performance of CNNs by minimizing data transfers
    from DRAM-based main memory. However, in these architectures, DRAM
    is still responsible for half of the overall energy consumption of the
    system, on average. A key factor of the high energy consumption of DRAM
    is the \emph{refresh overhead}, which is estimated to consume 40\% of the total DRAM energy.

    In this paper, we propose a new mechanism, \emph{Refresh Triggered
    Computation (RTC)}, that exploits the memory access patterns of CNN
    applications to reduce the number of \emph{refresh operations}. RTC
    uses two major techniques to mitigate the refresh overhead. First,
    \emph{Refresh Triggered Transfer (RTT)} is based on our \emph{new}
    observation that a CNN application accesses a large portion of the DRAM
    in a predictable and recurring manner. Thus, the read/write accesses
    of the application inherently refresh the DRAM, and therefore a significant fraction of refresh
    operations can be skipped. Second, \emph{Partial Array Auto-Refresh
    (PAAR)} eliminates the refresh operations to DRAM regions that do not
    store any data.
     
    We propose three RTC designs (min-RTC, mid-RTC, and full-RTC), each of which requires a different level
    of aggressiveness in terms of customization to the DRAM subsystem. All
    of our designs have small overhead. Even the most
    aggressive RTC design (i.e., full-RTC) imposes an area overhead of only 0.18\%
    in a $16\,Gb$ DRAM chip and can have less overhead for denser chips.
    Our experimental evaluation on six well-known CNNs show that RTC
    reduces average DRAM energy consumption by 24.4\% and 61.3\%, for the
    least aggressive and the most aggressive RTC implementations,
    respectively. Besides CNNs, we also evaluate our RTC mechanism on three
    workloads from other domains. We show that RTC saves 31.9\% and 16.9\%
    DRAM energy for \emph{Face Recognition} and \emph{Bayesian Confidence
    Propagation Neural Network (BCPNN)}, respectively. We believe RTC can
    be applied to other applications whose memory access patterns remain predictable for a sufficiently long time.

\end{abstract}

%\begin{CCSXML}
%<ccs2012>
%<concept>
%<concept_id>10010520.10010575.10010580</concept_id>
%<concept_desc>Computer systems organization~Processors and memory architectures</concept_desc>
%<concept_significance>500</concept_significance>
%</concept>
%<concept>
%<concept_id>10010583.10010600.10010607.10010608</concept_id>
%<concept_desc>Hardware~Dynamic memory</concept_desc>
%<concept_significance>500</concept_significance>
%</concept>
%</ccs2012>
%\end{CCSXML}
%
%\ccsdesc[500]{Computer systems organization~Processors and memory architectures}
%\ccsdesc[500]{Hardware~Dynamic memory}

%\keywords{DRAM, DRAM Refresh Overhead, Convolution Neural Networks}

\maketitle

%\setstretch{0.955}
\section{Introduction}
\label{sec:intro}

Neural Networks (NNs) are becoming a critically important class of
mainstream machine learning algorithms, as they provide high prediction
accuracy and are easily parallelizable~\cite{hinton2012imagenet,
chen2014diannao}. However, such benefits come at the cost of high
computational power and intensive memory usage, which require high energy
consumption. Convolutional Neural Networks (CNNs), a widely used type of
NNs, try to reduce computation and memory usage by sharing \emph{synaptic
weights} in each layer of the neural network.  Despite their relatively
efficient design, CNNs still require a significant amount of energy.
Furthermore, to process the information that is continuously received from various
sensors, emerging autonomous systems, e.g., self-driving vehicles, typically require multiple
simultaneously operating CNNs, which makes the energy consumed by CNNs even
more important. Hence, achieving low-power CNN implementations remains as a
challenging task.

As DRAM-based memory provides high capacity with decent latency, it is
typically used as main memory in systems that implement CNNs. Although DRAM
achieves high density by storing a single bit of data in the form of charge
in a DRAM cell, data stored in DRAM is volatile due to charge leakage from
the cell. To ensure data integrity, the charge of a cell needs to be
periodically replenished by refresh operations. DRAM refresh consumes
significant amount of energy and its overhead is expected to further
increase in future DRAM devices as DRAM capacity
increases~\cite{liu2012raidr, venkatesan2006retention, ghosh2007smart,
chang2014improving, refreshnow, mukundan2013understanding, reaper,
mutlu2013memory, mutlu2015research, chou2015reducing, khan2016parbor,
khan2017detecting, khan-sigmetrics2014, patel2020bit,
patel2019understanding, kang-memcon2014, liu2013experimental,
mutlu2017rowhammer, mutlu2019rowhammer, nair2013archshield, nair2013case,
nair2014refresh, nair2016xed}. For example, Liu et al.~\cite{liu2012raidr}
show that a single $4\,Gb$ DDR3 DRAM chip spends 15\% of the total DRAM
energy for refresh operations and project refreshes to consume
approximately half of the total DRAM energy in future $64\,Gb$ DRAM chips.
Thus, a DRAM device spends significant amount of energy only to ensure data
is stored correctly, even during idle periods where no DRAM accesses occur.

\begin{figure}[htbp]
    \begin{center}
		\includegraphics[width=.74\linewidth]{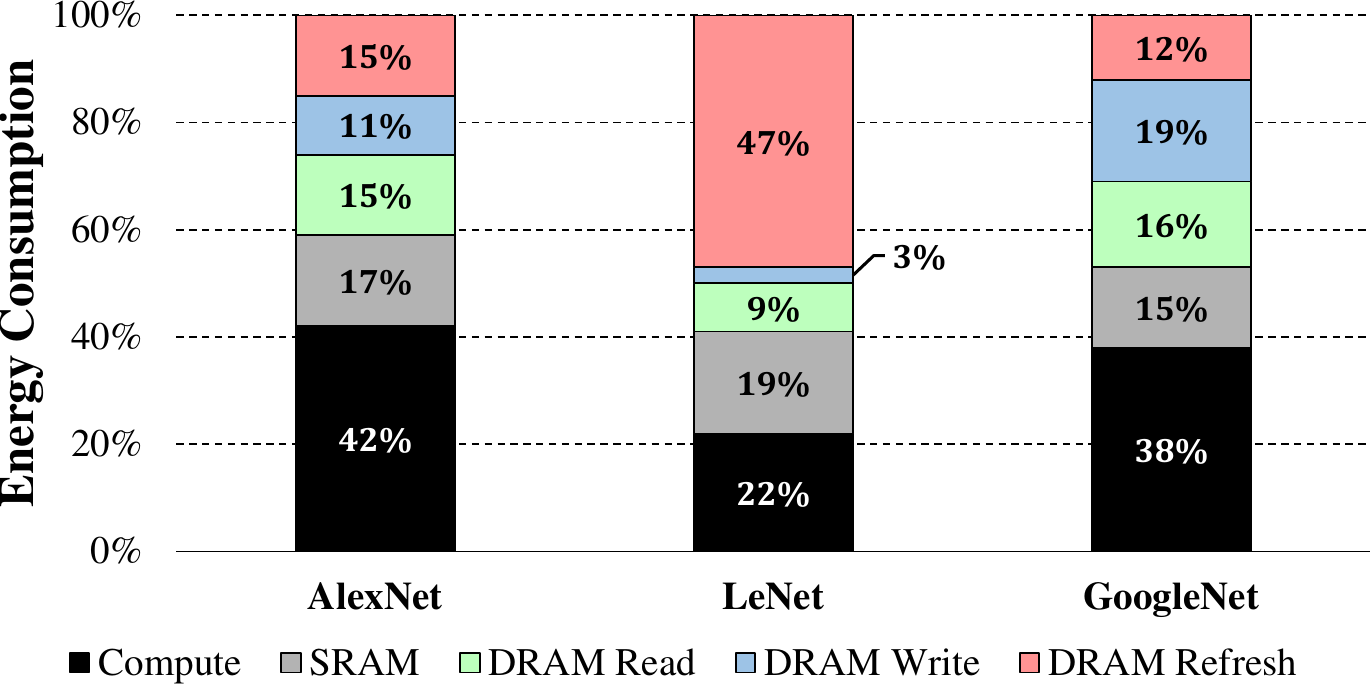}
	\end{center}
		\caption{Energy consumption breakdown of three CNNs on a modern CNN accelerator}
	\label{fig:cnn_energy_breakdown}
\end{figure}

CNNs typically have a large memory footprint~\cite{chen2016eyeriss}, mainly
due to a large number of synaptic weights that they maintain. Storing and
accessing the synaptic weights from the DRAM constitute the dominant
portion of energy consumption in CNNs~\cite{chen2016eyeriss}. To tackle this problem,
recently-proposed accelerators focus on reducing the DRAM accesses by
exploiting data locality~\cite{chen2014diannao, chen2016eyeriss, shi2015a, dally2016eie, mocha}.
Another approach compresses in-memory data to reduce the memory footprint
and data transfer overheads of CNNs~\cite{dally2016eie}. Although these
approaches improve energy consumption by reducing DRAM accesses, a CNN accelerator
still suffers from high DRAM refresh overhead.
Figure~\ref{fig:cnn_energy_breakdown} shows the energy breakdown of three
well-known CNNs, AlexNet~\cite{hinton2012imagenet},
LeNet~\cite{lecun1998gradient}, and GoogleNet~\cite{szegedy2015going},
which are implemented on an architecture similar to the state-of-the-art
Eyeriss~\cite{chen2016eyeriss} CNN accelerator.\footnote{Our methodology
and accelerator architecture are described in
Section~\ref{sec:rtc_concepts}.} The figure shows that the DRAM refresh
overhead constitutes a portion as large as 15\% for AlexNet and GoogleNet,
which are examples of large CNNs, and 47\% for LeNet, which is a relatively
smaller CNN. For these evaluations, we assume 2\,GB total DRAM capacity. For
higher capacity DRAM, which is common in systems today, the refresh
overhead is responsible for even larger portions of the overall DRAM energy
consumption~\cite{liu2012raidr} (see Section~\ref{sec:sensitivity_dram_density}). Thus, it
is critical to investigate and develop techniques that reduce the DRAM refresh
overhead for implementing energy-efficient CNNs.

Various mechanisms have been proposed to mitigate the DRAM refresh
overhead. Du et al.~\cite{du2015shidiannao} eliminate the refresh overhead
by implementing a CNN accelerator using only SRAM-based memory. Such an
approach not only restricts the applicability of the accelerator to small
CNNs, as a majority of CNNs typically require significant memory
capacity~\cite{dally2016eie, chen2016eyeriss, shi2015a}, but also increases
the energy consumption for storing synaptic weights as SRAM has higher
leakage power compared to DRAM with the same capacity.
Smart Refresh~\cite{ghosh2007smart} can reduce the refresh
overhead by skipping the refresh operation for a recently-accessed row.
However, to keep track of the time when a row was last accessed,
SmartRefresh introduces additional storage overhead by employing a counter
for each row. With the increase in DRAM capacity, the total storage
required by the counters exceeds one megabyte, overshadowing the energy
savings by reducing the number of refresh operations, as shown in ~\cite{liu2012raidr}.
There are also other works~\cite{agrawal2013refrint, pourshirazi2016refree, liu2012raidr, qureshi-dsn2015}
that propose mechanisms to reduce the DRAM refresh energy overhead.
However, they have high implementation cost or limited applicability, as
they require additional storage or can be applied only to embedded
DRAM~\cite{agrawal2013refrint}.

\textbf{Our goal} is to reduce the DRAM refresh overhead by eliminating the
unnecessary refresh operations with minimal overhead in CNN accelerators.
To achieve this, we propose a new technique that we call \emph{Refresh
Triggered Computation (RTC)}.  In RTC, we take advantage of two \emph{new}
observations to develop two orthogonal mechanisms for reducing DRAM energy
consumption by eliminating unnecessary refresh operations. First, we
observe that large CNNs, such as AlexNet~\cite{hinton2012imagenet}, access
DRAM periodically, \emph{with a fixed pattern}. As a read or write access
implicitly refreshes the accessed DRAM cells, we can exploit the access
pattern of such CNNs to overlap and replace read/write operations with the
refresh operations. To this end, we propose and implement a new
\emph{Refresh Triggered Transfer (RTT)} mechanism to coalesce the
read/write accesses with refresh operations. Second, we observe that
smaller CNNs, such as LeNet, leave most of the DRAM capacity unused. We
propose and implement \emph{Partial-array Auto Refresh (PAAR)}, which
eliminates the refresh operations to the portions of DRAM that are not
used. We find that large CNNs typically benefit more from RTT than from
PAAR, while the opposite is typically true for small CNNs that leave a
large portion of DRAM unallocated.

In this work, we implement and evaluate three variants of RTC that differ
in the level of customization required on the DRAM device and the memory
controller. The first variant, min-RTC, requires changes only in the memory
controller, and is useful when the read/write requests are frequent, such
that they can be coalesced with the refresh operations. For the second
variant, mid-RTC, we slightly modify the implementation of the
already-available \emph{Partial-array Self Refresh (PASR)} feature in
modern DRAM chips~\cite{micron_lpddr3, jedec-lpddr4}, to enable that
feature not only in \emph{self-refresh mode}, but also during normal
operation of the DRAM. For the third variant, full-RTC, we propose internal
DRAM modifications that fully exploit the capabilities of RTC. In
particular, we add an Address Generation Unit and a Finite State Machine
(FSM) to skip refreshes of recently accessed rows. In our evaluations, we
find that RTC reduces the DRAM refresh energy by 25\% to 96\% across six
different CNNs, depending on the used RTC variant, DRAM capacity, and the
access pattern of the application.

Although we apply RTC to mainly CNNs in the scope of this paper, a wide
class of applications with a \emph{pseudo-stationary spatio-temporal memory
access pattern} can take advantage of the RTC mechanism.
RTC reduces DRAM refresh overhead when the memory access
pattern of a workload is stationary for a time interval
sufficiently long enough to reconfigure the RTC
logic, and otherwise remains inactive with negligible system performance
and energy overhead. We believe that such long intervals with
stationary memory accesses are prevalent for a wide variety of streaming
applications (e.g., pattern recognition, signal processing,
computer vision) that operate on large amounts of data. We
demonstrate that multiple other applications, i.e., Face Recognition
and Bayesian Confidence Propagation Neural Network (BCPNN), significantly
benefit from RTC (Section~\ref{sec:other_applications}). We hope that
future work finds other use cases for RTC.

We make the following major contributions:

\squishlist

\item We observe that the regular memory access patterns of CNNs can be exploited
    to reduce the DRAM refresh overhead by replacing periodic refresh
    operations with read and write accesses.

\item We propose Refresh Triggered Computation (RTC) as a general technique
    to reduce the number of refresh operations based on applications memory
    access patterns. RTC includes two mechanisms: Refresh Triggered
    Transfer (RTT) for coalescing the read/write accesses with refresh
    operations, and Partial-array Auto Refresh (PAAR) for eliminating refreshes to portions of DRAM that are not being used.

\item To improve the adoption of RTC, we implement three variants of it
    that differ in the amount of modifications required to the DRAM device
    and the memory controller. We evaluate refresh overhead reduction of
    all three variants for six widely used CNN applications (i.e.,
    AlexNet~\cite{hinton2012imagenet}, LeNet~\cite{lecun1998gradient},
    GoogleNet~\cite{szegedy2015going}, Winograd~\cite{lavin2016fast},
    ResNet~\cite{he2016deep}, and Generative Adversarial
    Network~\cite{goodfellow2014generative}). We show that RTC, in its
    most aggressive variant, reduces DRAM refresh energy in a
    state-of-the-art CNN accelerator by up to 96\% (on average 61.3\%
    across multiple CNNs). We show that RTC is also effective for
    Face Recognition and Bayesian Confidence Propagation Neural Network
    (BCPNN) applications.

\squishend

\section{Background}
\label{sec:background}

In this section, we provide background on DRAM and CNNs, necessary to
understand the RTC framework that we propose. We refer the reader to past
works in DRAM for more details~\cite{keeth2007dram, jacob2010memory,
itoh2013vlsi, hassan2019crow, seshadri2017ambit, seshadri2019dram,
chang2017understanding, chang2016understanding, lee-hpca2015,
zhang2014half, ipek2008self, chang2014improving, lee2015decoupled,
lee2015simultaneous, chang2016low, luo2020clr, seshadri2013rowclone,
lee2017design, lee-hpca2013, hassan2017softmc, kim2018dram, kim2019d,
hassan2016chargecache, kim2018solar, ghose2019demystifying, kim-isca2012,
wang2020figaro, ghose2018your, kim2014flipping, kim2020revisiting,
wang2018reducing}.

\subsection{DRAM Organization and Operation}
\label{subsec:dram_background}
Dynamic Random Access Memory (DRAM) offers high memory density at
relatively low latency, which makes it the most preferable alternative for
implementing main memory on mobile, desktop, and warehouse-scale systems.
DRAM is also a viable option for CNN accelerators, as it
provides enough capacity to fit large CNNs.

\begin{figure}[htbp]
    \centering
    \includegraphics[width=.6\linewidth]{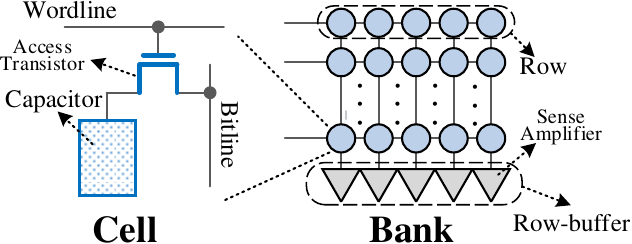}
    \caption{DRAM cell and bank}
    \vspace{-3mm}
    \label{fig:dram_arch}
\end{figure}

DRAM stores data in a hierarchical structure, as we show in
Figure~\ref{fig:dram_arch}. As the smallest component of the hierarchy, a DRAM
\emph{cell} stores a single bit of data in a \emph{capacitor} that is
accessed by enabling the \emph{access transistor} of the cell. As the cell
capacitor leaks its charge over time, to correctly maintain the data, the
capacitor needs to be \emph{periodically refreshed}, commonly once every $64 ms$.
Typically 2K to 16K cells are organized as a \emph{row}, where all cells
share the same \emph{wordline} connected to their access transistors.
Therefore, all cells in a row are refreshed simultaneously. The refresh
operation involves the \emph{sense amplifiers}, which are units that
connect to the cells via \emph{bitlines} and read the data out of the
corresponding cells based on the charge amount their capacitors store,
and correspondingly replenish the capacitor charge afterwards. As the area of a sense
amplifier is much higher than that of a DRAM cell~\cite{lee-hpca2013}, a large number
of cells from different rows share the same sense amplifier to provide
high memory density. However, as having an extremely large number of rows
that share a sense amplifier would negatively affect the access latency due
to increased parasitic capacitance on the bitline, the rows are grouped
into multiple \emph{banks}, where each bank has its own set of sense
amplifiers, referred to as \emph{row-buffer}. Besides improving access
latency, a banked structure also improves the memory throughput by
providing parallelism at bank-level (i.e., multiple banks can operate
simultaneously as they have separate row-buffers). Finally, at the top
level of the hierarchy, multiple chips are organized as a \emph{rank},
where the chips operate in lock step (i.e., perform the same operation
concurrently). There might be one or more ranks per
\emph{channel}. In the latter case, multiple ranks share the same memory
bus to interface with the processor, reducing I/O pin requirements, but limiting
parallelism.

DRAM commands are the interface between the memory controller and DRAM.
There are four main DRAM commands involved in a DRAM access. First, to
service a \emph{demand request} (i.e., a load or store request) the
memory controller issues an \emph{Activate (ACT)} command to select a row
from a bank, and copy its data to the row-buffer. After completion of that
operation, the memory controller can issue multiple \emph{READ} and
\emph{WRITE} commands to access the data in the row-buffer at a granularity
equal to the data bus width of the DRAM chip. In order to access data from
another row in the same bank, the memory controller first closes the
currently active row by issuing a \emph{Precharge (PRE)} command.

In addition to these four commands used to access DRAM, the memory
controller also periodically issues a \emph{Refresh (REF)} command to
replenish the charge stored in DRAM cell capacitors and ensure data
integrity. For the chips available in the market today, the entire DRAM
chip has to be refreshed every $64\,ms$~\cite{standard2007ddr3} (or
$32\,ms$ when operating at temperatures exceeding
$\SI{85}{\degreeCelsius}$~\cite{standard2007ddr3}). As there is a large
number of rows in the chip, the memory controller issues a refresh command
once every $7.8\,us$ to complete the refresh cycle for the entire DRAM in
$64\,ms$. A single refresh command typically refreshes multiple rows in
batch in hundreds of nanoseconds.\footnote{For an 8\,Gb DDR3 chip, a DRAM
refresh command takes 350\,ns to complete, during which all banks are
unavailable for access~\cite{standard2007ddr3,
chang2014improving}.} DRAM refresh consumes significant
amount of energy and its overhead is expected to further increase in future
DRAM devices as DRAM capacity increases~\cite{liu2012raidr,
venkatesan2006retention, ghosh2007smart, chang2014improving, refreshnow,
mukundan2013understanding, reaper, mutlu2013memory, mutlu2015research,
chou2015reducing, khan2016parbor, khan2017detecting, khan-sigmetrics2014,
patel2020bit, patel2019understanding, kang-memcon2014,
liu2013experimental}. For example, Liu et
al.~\cite{liu2012raidr} show that refreshes constitute 15\% of the total
DRAM energy for a $4\,Gb$ DDR3 chip and the fraction of DRAM energy spent
on DRAM refresh is projected to increase as DRAM chips become denser (e.g.,
refreshes would consume about 50\% of the total DRAM energy in future
$64\,Gb$ DRAM chips). Additionally, although a DRAM device may not be
always accessed with maximum throughput while executing a workload, all
DRAM rows have to be refreshed at a constant rate. Thus, when DRAM is
accessed infrequently, energy spent on DRAM refresh accounts for a
significant portion of the overall DRAM energy. We observe that typical CNN
workloads access DRAM regularly but not frequently enough such that refresh
operations consume significant DRAM energy compared to the energy consumed
by DRAM accesses.

An ACT-PRE command pair, which the memory
controller issues to service a demand request, also fundamentally performs the same operation as refresh. Both, first transfer
the charge stored in the capacitor to the sense amplifier, which later
fully restores the capacitor back to its original level (i.e.,
fully-charged or empty). As a result, both refresh and demand requests have the
ability to replenish the charge stored in the DRAM cells. We exploit this observation in the design of our mechanism to save DRAM refresh energy.

\subsection{Convolutional Neural Networks}
\label{subsec:conv_nns}

Convolutional Neural Networks (CNNs)~\cite{lecun1998gradient} are machine
learning algorithms that achieve state-of-the-art learning accuracy. The
basic idea of CNNs is to extract low-level features from the input data at
high resolution, and later combine those features to build more complex
ones.  

\begin{figure}[htbp]
    \centering
    \includegraphics[width=\linewidth]{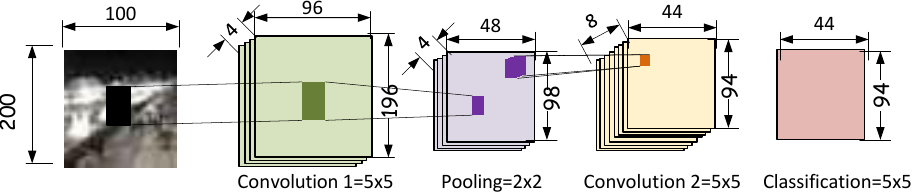}
    \caption{The general structure of a CNN}
    \vspace{-3mm}
    \label{fig:convexample}
\end{figure}

As we show in Figure~\ref{fig:convexample}, a CNN consists of multiple layers,
which contain feature maps at different abstraction levels of the input
data, and synaptic weights (i.e., convolutional kernels), which are used
for extracting the features of the next layer by performing convolution on
the output of the previous layer. There are two main computational phases in a CNN:
\emph{training} and \emph{inference}. During the training phase, to learn
what to infer from the input data, the CNN processes a large amount of
reference data using error back-propagation~\cite{rumelhart1985learning}.
Later, during the inference phase, the CNN classifies the input data by
using the information that it has learned during the training phase. In
general, it is sufficient to perform the training phase offline, before the
inference phase~\cite{chen2014diannao, shi2015a}. Since the offline
training does not affect the performance of the end-application, we
focus on the inference phase, similar to prior work~\cite{dally2016eie,
chen2016eyeriss, shi2015a, chen2014diannao}. However, we observe that the
training phase exhibits similar memory access patterns as the inference
phase, and thus the techniques we propose can also be applied to
the training phase.

The high-level goal of the inference phase is to infer the required
information (e.g., whether a particular object is available in the image)
from raw input. CNNs have multiple layers between the input data and
the final classification output. Primarily, there are three types of
layers: (i) \emph{convolution}, (ii) \emph{pooling}, and (iii)
\emph{classification} layers. The convolutional layer extracts various
features (e.g., edges and corners) by convolving a 2D mask (of synaptic
weights) with the input data from the previous layer. For each
feature that is being extracted, the CNN applies a different set of synaptic weights to
the input, producing multiple feature maps. For example, in
Figure~\ref{fig:convexample}, the first convolutional layer (Conv1)
produces four output feature maps by convolving the input image with a 5x5
mask. The pooling layer extracts the salient features from the previous
layer, usually by applying a max or averaging function. After several layers
of convolution and pooling, the input image is classified in the
classification layer, which provides the probability that the input belongs
to a particular class.

The inference phase of the CNN is largely memory intensive.
When processing an input image, the CNN needs to read the large data (i.e.,
synaptic weights and outputs of previous layer) of each layer
from the memory. For each layer, the CNN runs multiple convolution or
pooling operations and writes back the results to memory. Thus, the
inference phase yields a large read and write traffic that could not be
entirely filtered out by the caches, and requires the data to be serviced
from DRAM. For example, AlexNet~\cite{hinton2012imagenet} performs
about 3 billion DRAM accesses when processing a single image. Modern CNN
accelerators~\cite{chen2016eyeriss} reduce this requirement to 60 million
DRAM accesses per input image by exploiting data locality. However, despite
that huge reduction, DRAM is still major contributor to the overall
energy consumption of a system, as we see in Figure~\ref{fig:cnn_energy_breakdown}.

%\setstretch{0.93}
\section{Refresh Triggered Computation}
\label{sec:rtc_concepts}

As we explained in Section~\ref{subsec:dram_background}, the memory controller
periodically issues refresh commands to DRAM, in order to ensure data
integrity. 
Such frequent and time-consuming refresh operations
often conflict with read and write requests that are issued by the
workloads running on the system~\cite{chang2014improving, liu2012raidr}. As a result, a
refresh operation not only consumes significant amount of energy, but
also negatively affects system performance by delaying read
and write requests.

\emph{Refresh Triggered Computation (RTC)} is based on the high-level observation that
for applications with regular memory access patterns, such as CNNs, it is possible to synchronize
the refresh operations and the read/write requests, such that read/write requests of the application naturally refresh DRAM.
RTC not only eliminates conflicts between refresh and read/write, but also
reduces the number of refresh commands that the memory controller needs
to issue by eliminating redundant refresh operations. 
In this section, we first introduce the RTC concepts before elaborating on RTC's
implementation details in Section~\ref{sec:architecture}.

\subsection{Making Refresh Unnecessary}

% remind that read/write accesses also replenish the row's charge as the
% refresh operation does
In Section~\ref{subsec:dram_background}, we explain that both refresh and
access requests perform similar operations (i.e., activating and precharging a row) in the DRAM circuitry that replenish the charge of the DRAM cells in a row. We observe that, in many cases, the \textit{explicit} refresh operations can be eliminated, since i) the DRAM access requests are at least as frequent as the periodic refresh operations and ii) such requests continuously cover a very large portion of the DRAM. Hence, there is potential to eliminate most of the \textit{explicit} refresh operations since a large fraction of the DRAM is already being \textit{implicitly} refreshed when accessed.

We aim to make refresh unnecessary by ensuring that the row to be refreshed is accessed at the same time it is supposed to be refreshed. 
However, performing such an alignment is not straightforward due to two
reasons. First, the periodic refresh operation is performed using an
in-DRAM counter that points to the next row to be refreshed. 
Thus, an application (or even the memory controller) does \textit{not} have
control on \textit{which} row will be refreshed next. Second, the access
requests are \textit{not} as regular as the refresh operations, in
terms of their row access pattern. Therefore, aligning refresh
operations with accesses is a challenging problem.

\subsection{Alignment in a Controlled Environment}
\label{subsec:simple_alignment}

% explain the technique designed to align refresh and read/write, RTT
To develop a feasible and efficient solution for the problem of aligning
the access requests with the periodic refresh operations, we first make
three simplifying assumptions. \emph{i)} We assume that the access pattern
of the application is known in advance and it is periodic. In other words,
the application has an iterative execution flow and, in each iteration, it
generates requests in a fixed order. \emph{ii)} The period of the
access requests is lower than (or same as) the period of the
refreshes. This assumption ensures that the refresh period (e.g., 64ms) of
a DRAM row is not exceeded between two consecutive accesses to the row.
\emph{iii)} We assume that the entire working data set of the application
is accessed in each iteration. In Sections~\ref{subsec:rtt}
and~\ref{subsec:paar}, we introduce our techniques to handle the cases
where these assumptions do \textit{not} hold true.

\begin{figure}[htbp]
    \centering
        \includegraphics[width=.9\linewidth]{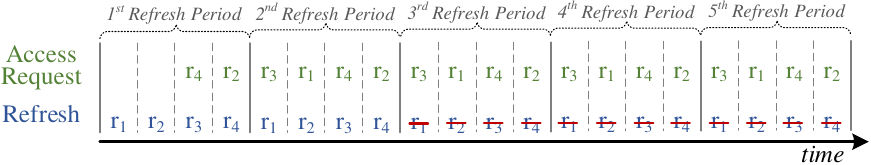}
        \vspace{-2mm}
        \caption{Periodic application access pattern vs. refresh pattern.}
		\vspace{-4mm}
    \label{fig:aligning_refresh}
\end{figure}

% aligning refresh and demand accesses when the three assumptions hold
In the process of making refreshes unnecessary, we first design a scheme
that aligns refreshes with reads when the three assumptions about
the applications access pattern hold. In Figure~\ref{fig:aligning_refresh},
we explain how such a scheme works by plotting a timeline of accesses
that an application performs and refreshes that the memory controller
issues during three refresh periods. The refresh requests iterate through
rows $r_1$ to $r_4$ in the first two refresh periods. Close to the end of
the first refresh period, the application starts to issue access requests
to all of these four rows, but in different order. In the second period,
the refresh operations are still required to ensure data integrity
because if we eliminate the refresh operations, the time since $r_1$
was last refreshed would exceed the refresh period. In contrast, all
refresh operations in the third period are redundant as the rows are
already refreshed due to the accesses in the same period. Next, we
introduce our techniques to align refreshes and access request when the
three simplifying assumptions are relaxed.

\subsection{Refresh Triggered Transfer}
\label{subsec:rtt}

The simple scheme we proposed in Section~\ref{subsec:simple_alignment}
assumes that the rate of the refresh operations and accesses
always match. However, in real applications, the access requests can be
more or less frequent than the refreshes.

To solve the problem of matching the rates of accesses and refreshes, we
propose \emph{Refresh Triggered Transfer (RTT)}. The key idea of RTT is
to alter the existing periodic refresh scheme to align the
refreshes with the access requests.  We achieve this by slightly
modifying the DRAM auto-refresh circuit, as we explain in
Section~\ref{subsec:full_rtc}.

Algorithm~\ref{alg:rate_matching_algo} describes how RTT handles the
mismatch in the rates of accesses and refreshes.\footnote{We
adapt the algorithm from a technique~\cite{jeanvtvlsi2014,
asadisqed2012} that is used to align send and receive processes operating
at rationally related clock frequencies.} If the application generates
access requests to its entire allocated memory as
frequently as the refresh rate or faster, RTT completely removes the
refresh overhead and ensures data integrity as an access (i.e., DRAM
row activation and precharge) replenishes the charge of a row it
accesses. However, when the accesses are not frequent enough, the
problem of matching the rates of the accesses and refreshes becomes
more challenging. To tackle this problem, RTT eliminates refreshes
\emph{partially} by performing refresh only on rows that are \emph{not}
accessed within the refresh period.

Algorithm~\ref{alg:rate_matching_algo} takes $N_a$ and $N_r$ as
input\footnote{$N_r$ is equal to the number of rows in DRAM, as the entire
DRAM needs to be refreshed in a single refresh period.}, which are the
number of rows that the access requests and refreshes target during a
single refresh period, respectively. The output of the algorithm is the
explicit refresh ($exp\_ref$) signal, which determines whether a row will
be explicitly refreshed or implicitly replenished when accessed to
read/write data. Thus, when $N_r \leq N_a$, $exp\_ref$ is set as 0 to
indicate that an access occurs to all rows frequently enough (line 4). When
the opposite is the case, i.e., $N_r > N_a$, then the algorithm needs to
output additional refresh operations to compensate for the rows that are
\emph{not accessed} during the refresh period. To find which rows to
refresh using explicit refresh requests, the algorithm starts with a credit
$c$, equal to $N_r$ (line 7). For each implicit refresh, $c$ is reduced by
$N_r - N_a$, until the credit becomes less than $N_r - N_a$. At this point,
the algorithm signals $exp\_ref = 1$ to indicate an explicit refresh, and
increments the credit by $N_a$.

\begin{algorithm}
\caption{Rate matching algorithm}
\label{alg:rate_matching_algo}
    \begin{algorithmic}[1]
    \Statex \(\triangleright\) $N_a$: \textit{the number of rows accessed by read/write during a refresh period}
    \Statex \(\triangleright\) $N_r$: \textit{the number of rows refreshed during a refresh period}
        
    \Procedure {RateMatching}{$N_a$, $N_r$}
        \For {every refresh period}
            \If {$N_r \leq N_a$}
                \State $exp\_ref \leftarrow 0$ \Comment implicit refresh
            \Else
                \State $P \leftarrow N_r/gcd(N_r, N_a)$
                \State $c \leftarrow N_r$
                \For {$i \leftarrow 1, P$}
                    \If {$c > N_r - N_a$}
                        \State $exp\_ref \leftarrow 0$ \Comment implicit refresh
                        \State $c \leftarrow c - (N_r - N_a)$
                    \Else
                        \State $exp\_ref \leftarrow 1$ \Comment explicit refresh
                        \State $c \leftarrow c + N_a$
                    \EndIf
                \EndFor
            \EndIf
        \EndFor
    \EndProcedure
        \end{algorithmic}
\end{algorithm}

\begin{figure}[htbp]
    \centering
        \includegraphics[width=.5\linewidth]{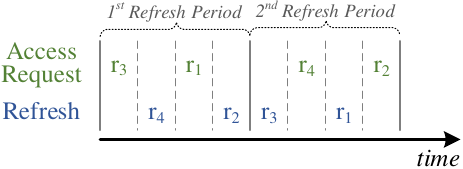}
        \vspace{-2mm}
        \caption{Accesses and refreshes generated by RTT for $N_r = 4$ and
        $N_a = 2$}
        \vspace{-3mm}
    \label{fig:rtt_example}
\end{figure}

To understand how Algorithm~\ref{alg:rate_matching_algo} operates, consider
an example where $N_a = 2$ and $N_r = 4$. Within a refresh period, only
half of the rows will be refreshed using an explicit refresh operation, as
we illustrate in Figure~\ref{fig:rtt_example}. Initially, $P=1$ and $c=4$
(lines 6-7). In the first iteration of the loop (line 8), $c$ is greater
than $N_r - N_a = 2$. Thus, the row is implicitly refreshed, and the credit
is decreased (lines 10-11). In the next iteration, as the credit is not
greater than $N_r - N_a$, an explicit refresh will be triggered (line 13).
Thus, the algorithm will interleave between an implicit and an explicit
refresh operation. We implement the RTT scheme in DRAM with minor
modifications to existing circuitry as we explain in
Section~\ref{sec:architecture}.

\textbf{Generating Memory Access Patterns.} The existing refresh
scheme implements a counter in the DRAM chip to refresh the rows
with a fixed pattern. However, the access pattern of a real
application may not follow the same pattern as the refreshes.
To adapt the refresh scheme to arbitrary access patterns, RTT implements
an \emph{Address Generation Unit (AGU)} that is similar to the
proposal in prior work~\cite{farahini2014parallel}. AGUs are commonly
used in Digital Signal Processors (DSPs) to efficiently generate the memory
addresses to feed to the functional units~\cite{liu2008embedded,
velilla2009scratchpad, farahini2014parallel, taniguchi2009systematic,
udayanarayanan2001address, leupers1996algorithms}. An AGU can typically be
programmed to generate various address sequences for a given application.
Prior works propose a broad range of AGU designs that can generate address
sequences with various amounts of complexity (e.g., commonly used DSP
addressing modes such as bit-reverse and circular buffer
addressing~\cite{instruments2002tms320c55x, xue2005optimizing}, piece-wise
affine address pattern generation~\cite{ilic2011address}, two-dimensional
affine address generation~\cite{mathew2004loop}, complex addressing modes
that include multiplication, modulo, and
shift~\cite{taniguchi2009reconfigurable}). We observe that the memory
access patterns of the workloads we focus on in the scope of this work are
relatively regular, and thus, to keep the design simple, we adopt an AGU
design that can generate address sequences based on an arbitrary affine
function. We explain the details of AGU's implementation in
Section~\ref{subsec:full_rtc}.

\subsection{Partial-Array Auto Refresh}
\label{subsec:paar}

For many applications, a significantly large portion of the DRAM may not
always be in use (i.e.,  portions may be unallocated). For example, the
memory footprint of LeNet~\cite{shi2015a}, which is a small CNN, is only
$1.06MB$ (e.g., when 100*100 image is used for character
recognition).  Hence, depending on the DRAM capacity, a large
number of unallocated DRAM rows would unnecessarily be refreshed,
consuming significant energy. In RTC, we implement a technique,
\emph{Partial-Array Auto Refresh (PAAR)}, which ensures that
refreshes are generated only for rows that are allocated.

PAAR should not be confused with a technique called \emph{Partial-Array
Self Refresh (PASR)}~\cite{jedec-lpddr4}, which already exists in low-power
DRAM chips and is used to refresh only certain DRAM banks while in
self-refresh mode (i.e., power-saving mode in which DRAM cannot be
accessed). As PASR operates at coarse bank granularity, no data should be
allocated in an entire bank that PASR turns refresh off. PAAR differs
from PASR mainly in two ways. First, to enable PASR, the memory
controller needs to switch the DRAM to a special low-power mode. Besides
switching in and out of this mode is a relatively slow
process~\cite{Mall-micro2012}, another downside of PASR is
that the DRAM cannot serve access requests while in PASR mode. In
contrast, PAAR can be enabled during the normal operation of the DRAM.
Second, PASR can eliminate refreshes only at bank-granularity. In order to
eliminate refreshes using such a scheme, an entire bank should be
unallocated.  Leaving one or more banks out of data allocation limits
bank-level parallelism~\cite{Mall-micro2012, mutlu2008parallelism}, and
reduces the memory bandwidth. In contrast, PAAR operates at row-granularity
and thus provides a more practical scheme to eliminate redundant refreshes
compared to PASR.

\subsection{Limitations of RTC}
\label{subsec:rtc_limitations}

Our RTC framework has two limitations.

\textbf{Access Patterns.} RTC can eliminate redundant refresh operations
when the access pattern of an application is stationary for sufficiently
long time. Configuring the AGU of RTC can take approximately 100
cycles. To compensate for this latency overhead, the access pattern
of the application should not change very frequently. Fortunately, there
are many applications from different domains (e.g., signal processing,
neural networks, bioinformatics) that exhibit regular access patterns.
In this work, we expect the programmer to determine the memory
access pattern of an application. However, a profiling-based or
compiler-assisted approach can potentially be used to automatically
determine access patterns of applications and take advantage of RTC without
involving the programmer. We leave this study to future work. For
other applications that have frequently changing access patterns, RTC can
be disabled to operate DRAM in the conventional way with negligible
performance and energy overhead.

\textbf{Simultaneously Running Applications.} Even though two different
applications have regular access patterns, running them simultaneously on
the same system may lead to irregularity in the memory access pattern. To
support multiple applications, we propose to map applications to separate
DRAM banks or channels, each with its own RTC control logic. Note
that such an approach does not reduce the bank-level parallelism,
since all banks continue to receive memory requests, but from
different applications. In fact, prior work shows that partitioning the
applications to separate banks or channels improves overall system
performance by reducing the bank/channel
conflicts~\cite{liu2012software, muralidhara2011reducing,
jeong2012balancing}.

%\setstretch{0.955}
\section{The RTC Architecture}
\label{sec:architecture}

In this section, we present the RTC architecture, which implements the
concepts we introduced in Section~\ref{sec:rtc_concepts}. We propose three
variants of RTC, differing in the level of customization that they require.
First, \emph{Min-RTC} does not require any changes to the DRAM
chip, but it only slightly changes the memory controller.
Second, besides the changes to the memory controller, \emph{Mid-RTC}
also introduces minimal modifications the DRAM peripheral logic.
Third, our most aggressive implementation, \emph{Full-RTC}, exploits
the full potential of the RTC concepts.

\subsection{Min-RTC}
\label{subsec:min_rtc}

For this implementation, we restrain ourselves from making any changes to
the DRAM chip. By modifying only the memory controller, we can implement
RTT partially and cannot implement PAAR at all. Thus, \emph{Min-RTC} is
only useful when the accesses are more frequent than the refreshes such
that all refresh operations can be eliminated.

With min-RTC, the memory controller receives information about the access
period directly from the application. Based on the information, the memory
controller decides whether to operate in \emph{normal} or \emph{min-RTC
mode}. If the application accesses the memory slower than the refresh rate,
the memory controller disables min-RTC, and operates in normal mode.
Otherwise, it enables min-RTC to eliminate the overhead of the refresh
operations. To achieve this, first, the memory controller aligns the
accesses with the refreshes as we describe in
Section~\ref{subsec:simple_alignment}. Later, the memory controller stops
issuing refresh commands to DRAM, as the access requests implicitly refresh
DRAM. The memory controller disables min-RTC when the application completes
execution or another application is invoked. According to our evaluations
(Section~\ref{sec:energy_cnns}), even such a simple mechanism saves
significant energy.

\subsection{Mid-RTC}
\label{subsec:mid_rtc}

In mid-RTC, besides the changes required for min-RTC, we also apply minor
modifications to the DRAM control logic to enable a coarse-grained
(bank-granularity) implementation of PAAR. Particularly, we modify
the logic that enables PASR, which is already available in low power DRAM
chips~\cite{lowpowerddr}, but is used only when the DRAM chip
is in low-power stand-by mode. To enable PAAR, we reuse the PASR logic and
make it possible to activate even when the DRAM is in normal mode of
operation. In mid-RTC, we avoid adding additional registers to define the
range of rows that will be refreshed with PAAR, and thus PAAR operates at
bank-granularity in this implementation.

Mid-RTC can mitigate the refresh overhead by eliminating
unnecessary refreshes, as min-RTC does, and by disabling the refreshes for
the DRAM banks that do not have any allocated portions.

\subsection{Full-RTC}
\label{subsec:full_rtc}

\begin{figure}[htbp]
    \centering
        \includegraphics[width=.9\linewidth]{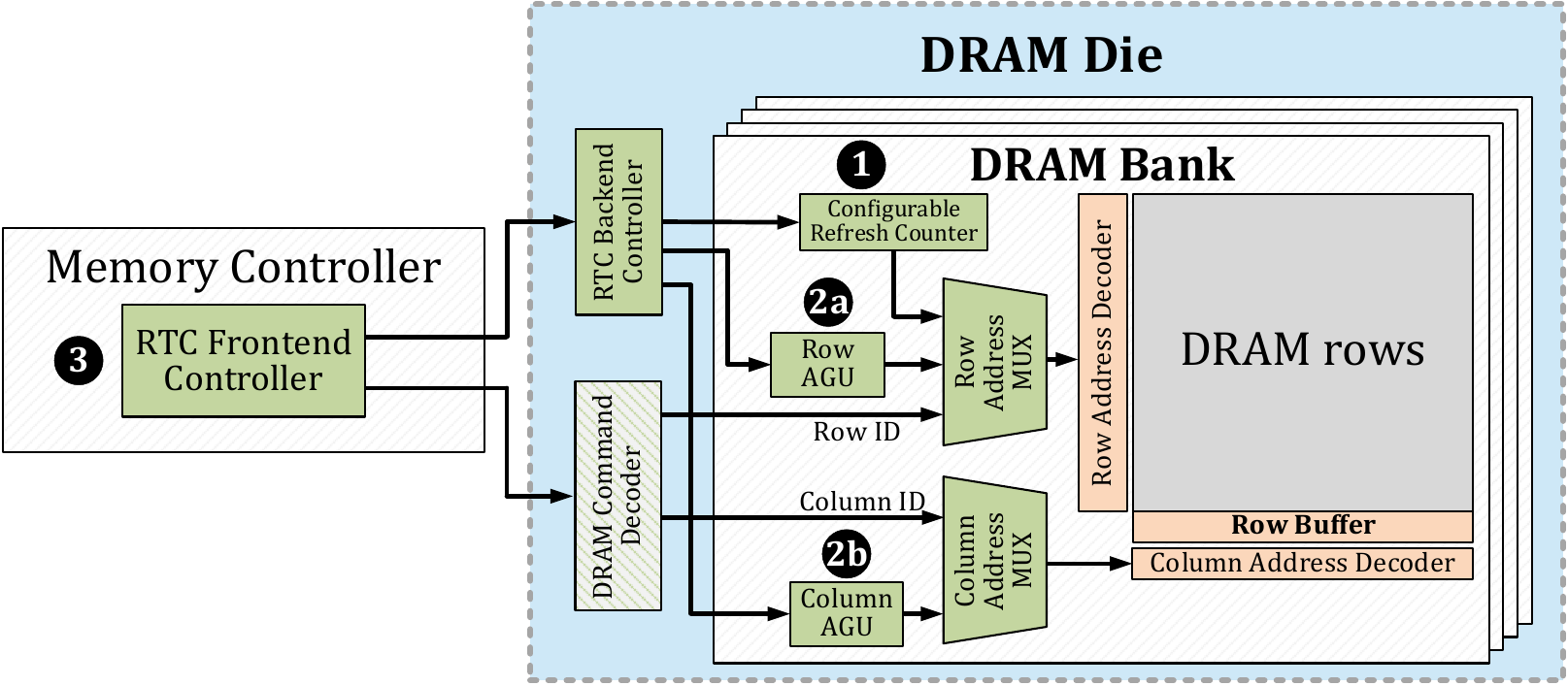}
        \caption{Modifications in the memory controller and DRAM to support Full-RTC}
        \label{fig:modifiedddr3}
\end{figure}

As we show in Figure~\ref{fig:modifiedddr3}, the most aggressive
implementation, full-RTC, requires mainly three modifications in
the DRAM chip and the memory controller. \circled{1} To prevent a subset
of non-allocated DRAM rows from being refreshed, full-RTC modifies
the in-DRAM refresh logic to be configurable by the memory controller.
\circled{2} To \emph{fully} implement the RTT scheme as
described in Section~\ref{subsec:rtt}, full-RTC adds an \emph{Address
Generation Unit (AGU)}, which is implemented in two levels (i.e., \emph{Row
AGU}~\circled{\small{2a}} and \emph{column AGU}~\circled{\small{2b}}).
An application can configure the AGU at runtime
to generate access and refresh requests using an arbitrary affine
function. \circled{3} Full-RTC implements \emph{RTC Frontend
Controller} to enable reconfiguration of the AGUs and the \emph{refresh
counter}; and executes Algorithm~\ref{alg:rate_matching_algo} to
determine which addresses generated by the AGU will transfer data
from/to DRAM, and which will only refresh the corresponding row.
Full-RTC implements this algorithm in the memory controller (i.e., in
the RTC Frontend Controller) but it also introduces a small
modification to the DRAM command decoder to handle the
explicit refresh ($exp\_ref$) signal generated by the RTC
Frontend Controller. We explain our design in more detail.

\subsubsection{Modifications to the Memory Controller}
\label{subsubsec:mctrl_changes}

The memory controller is the interface between the accelerator/processor
and DRAM. We modify the memory controller to support our changes in the
DRAM architecture that enables full-RTC.

In full-RTC, applications need to provide their memory access
patterns to the RTC Frontend Controller, which reconfigures the AGU
and the \emph{refresh counter}. Once the RTC Frontend Controller completes
reconfiguring the AGU, the AGU starts generating DRAM row and column
addresses to access data according to how the application has configured
the AGUs.

The address generation unit (AGU) incorporated inside the RTT counter logic
can be configured with an arbitrary affine function to generate
various memory access patterns that applications typically
exhibit. In our implementation, the memory controller uses special
commands to configure the AGU.

\subsubsection{Modifications to the DRAM chip}
\label{subsubsec:dram_changes}

PAAR improves DRAM energy efficiency by eliminating the refresh operations
to DRAM regions that are not allocated. In conventional DRAM,
periodic refresh operations are performed on all DRAM rows with a fixed
pattern. We slightly modify the conventional control logic for the
periodic refresh operations to limit the refreshed address range such
that only a specific region of DRAM is refreshed. As we illustrate in
Figure~\ref{fig:modifiedddr3}, we implement this feature by
introducing a \emph{Configurable Refresh Counter}, which incorporates
a register for start and end row addresses of the region to refresh. The
\emph{RTC Backend Controller} provides an interface to the memory
controller to configure start and end addresses of the DRAM rows to
refresh.

\subsubsection{RTC Controller Operation}
\label{subsubsec:rtc_controller}

The RTT technique aligns memory accesses with refresh requests such that
explicit refresh operations can be eliminated as accesses already
implicitly refresh DRAM rows. To achieve the alignment of
accesses and refreshes, the RTC Frontend Controller implements
Algorithm~\ref{alg:rate_matching_algo} that we explain in
Section~\ref{subsec:rtt}. By running the algorithm, the RTC Frontend
Controller determines whether DRAM should perform an access using the next
address generated by the AGU or a refresh operation using the refresh
counter.

\begin{figure}[htbp]
    \centering
        \includegraphics[width=0.7\linewidth]{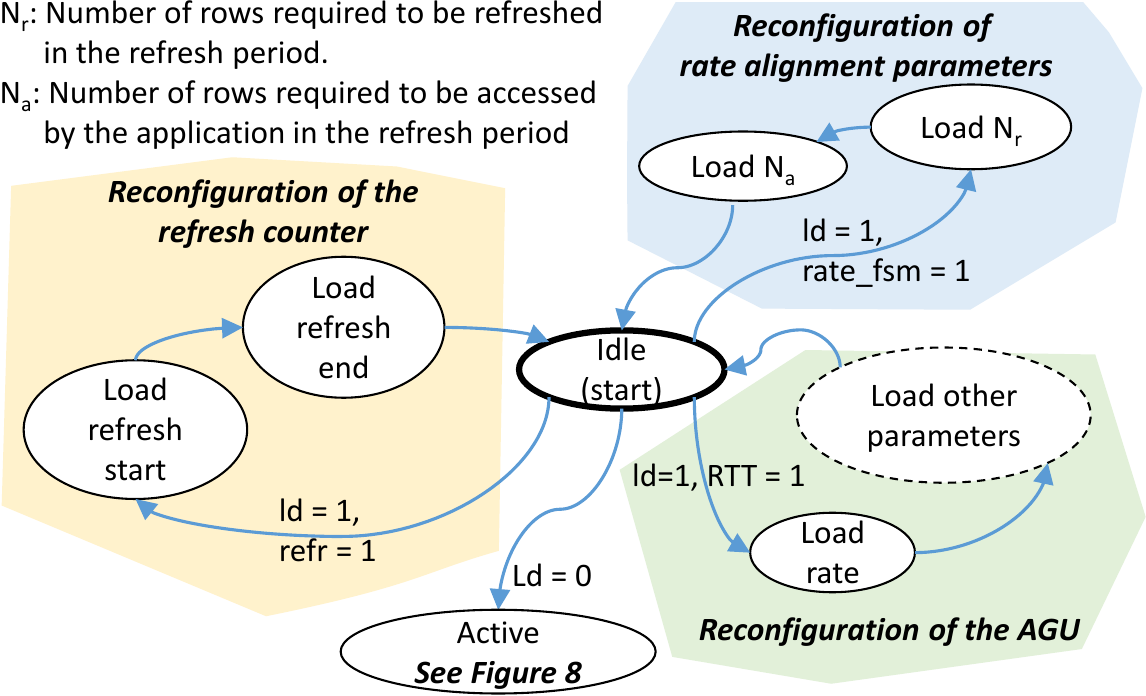}
        \caption{Operation of the RTC Frontend Controller}
    \label{fig:statemachineinitialization}
\end{figure}

\begin{figure}[htbp]
    \centering
        \includegraphics[width=0.7\linewidth]{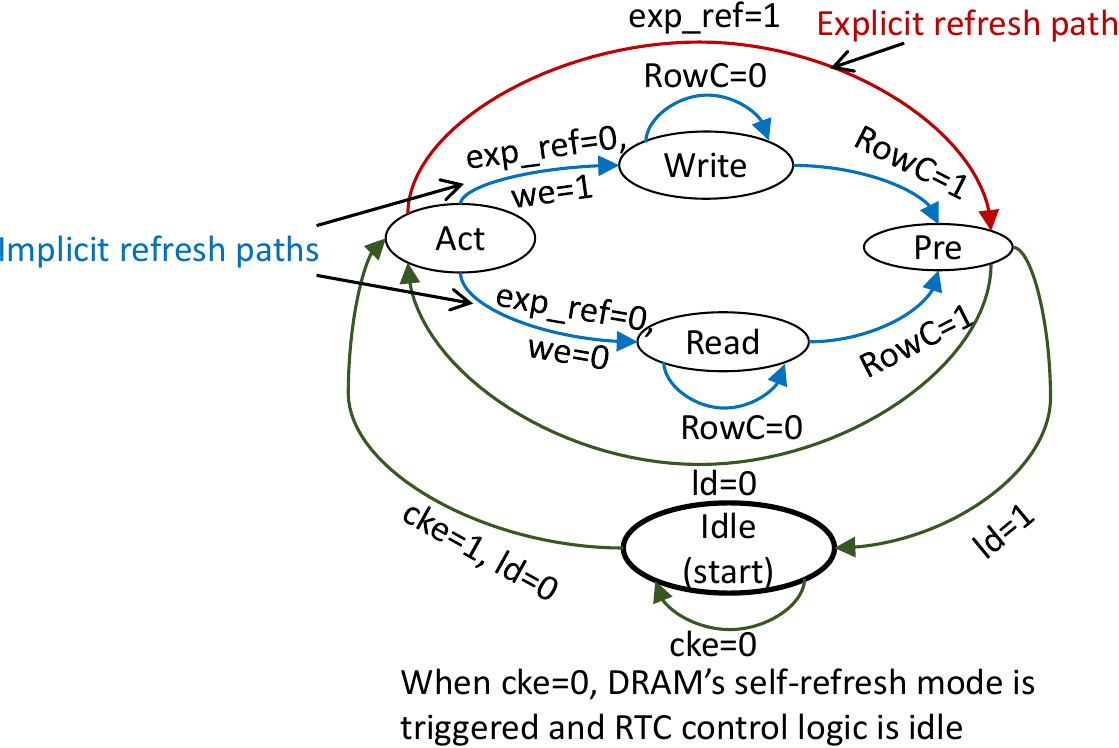}
        \caption{State machine that describes full-RTC operation}
    \label{fig:statemachinefullrttskiprtt}
\end{figure}

In Figure~\ref{fig:statemachineinitialization}, we describe the
operation of the \emph{RTC Frontend Controller} using a state
diagram. During the initial $idle$ state, the \emph{RTC Frontend
Controller} expects signals for reconfiguring one of its three components
(shaded with different colors). Once reconfigured, it transitions
into the $Active$ state, where RTT is enabled (we describe operation
in $Active$ state in Figure~\ref{fig:statemachinefullrttskiprtt}).

To enter a reconfiguration state, the $load$ signal ($ld$) has to be
asserted along with one of the three signals that indicate which of
the reconfiguration states to enter. First, when \texttt{refr=1},
the \emph{RTC Controller} reconfigures the start and end row
addresses of the \emph{Configurable Refresh Counter}. Second, when
\texttt{rtt=1}, the \emph{RTC Controller} reconfigures the Row and
Column AGUs. Third, when \texttt{rate\_fsm=1}, the \emph{RTC Controller}
reconfigures the $N_a$ and $N_r$ parameters that we describe in
Section~\ref{subsec:rtt}.

In Figure~\ref{fig:statemachinefullrttskiprtt}, we show a
diagram that describes the operation of RTT. While RTC
reconfiguration is in progress, the \emph{CKE} signal remains low to keep
RTT in \emph{idle} state. After reconfiguration finishes, the memory
controller starts RTT operation by setting \texttt{CKE} and \texttt{ld} to
0.

In the \emph{Act} state, the \emph{RTC Backend Controller} generates a DRAM
command either to activate the row at the address that the Row AGU provides
or to refresh the row that the \emph{Configurable Refresh Counter} points
to depending on the $exp\_ref$ signal sent by the memory controller. First,
if the time interval between two consecutive read/write requests is greater
than the required refresh interval (i.e., the memory controller sends
$exp\_ref=1$), the RTC Backend controller \emph{explicitly} refreshes the
row that the \emph{refresh counter} points to. We show this in
Figure~\ref{fig:statemachinefullrttskiprtt} with a red line from the
\emph{Act} to the \emph{Pre} state.  During this state transition, the
memory controller issues a precharge command to close the open row. Second,
when $exp\_ref=0$, control is transferred to either the \emph{Read} or the
\emph{Write} state depending on whether the write enable (\emph{we}) signal
is set to 1 or 0. This is because if the read/write path is taken, the rows
are implicitly refreshed. RTT remains operational as long as \texttt{ld=0}.
When \texttt{ld=1}, the control returns to the \emph{idle} state in
Figure~\ref{fig:statemachineinitialization}, which allows the RTC to be
reconfigured.

\section{Methodology}
\label{sec:methodology}

\begin{figure}[!hb]
    \centering
        \includegraphics[width=0.32\linewidth]{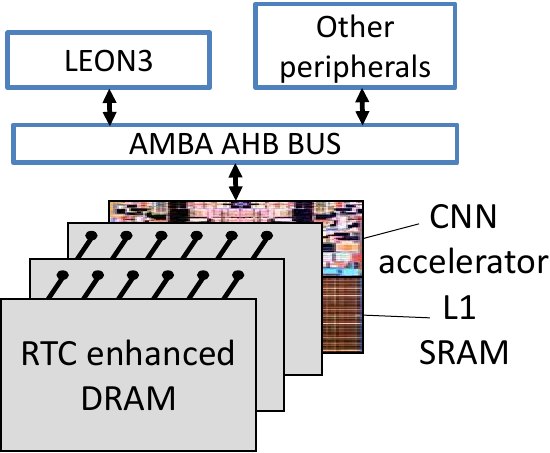}
        \caption{System-level view of the proposed architecture}
        \vspace{-3mm}
    \label{fig:systemoverview}
\end{figure}

We implement the RTC framework on a system that consists of a LEON3-based
open-source processor, which is connected to a state-of-the-art CNN
accelerator MOCHA~\cite{mocha}, similar to
Eyeriss~\cite{chen2016eyeriss}, via an AMBA AHB
bus~\cite{specification1999rev}. As we illustrate in
Figure~\ref{fig:systemoverview}, the accelerator is implemented in the
logic-layer of a DRAM-based 3D-stacked memory. We evaluate DRAM
capacities of $16\,Gb$, $32\,Gb$, and $64\,Gb$. The CNN
accelerator has a private 108KB scratch-pad memory, as in
Eyeriss~\cite{chen2016eyeriss}, and it also incorporates a memory
controller to interface the upper DRAM layers of the 3D-stacked memory.
The Eyeriss architecture uses row-stationary dataflow, which
aims to maximize reuse of filter weights and feature maps in the processing
engines' local storage to minimize DRAM accesses.

To analyze the effectiveness of RTC at saving DRAM refresh energy,
we evaluate six widely-used CNN applications,
GoogleNet~\cite{szegedy2015going}, AlexNet~\cite{hinton2012imagenet},
LeNet~\cite{lecun1998gradient}, Winograd~\cite{lavin2016fast},
ResNet~\cite{he2016deep}, and Generative Adversarial Network
(GAN)~\cite{goodfellow2014generative}. We adjust the batch
size for each CNN individually depending on how many kernels we can
accommodate at most in the register files of the MOCHA accelerator.

Winograd is an algorithm for efficiently performing convolution operations.
Winograd promises reduction in multiplication operations but it does not
affect main memory access characteristics. Reducing the number of
multiplications is beneficial for architectures such as GPUs and FPGAs that
perform convolutions as matrix multiplication. However, custom CNN
accelerator architectures already implement hardware optimizations to
perform convolution efficiently, and therefore using Winograd has
negligible impact on performance and energy consumption of CNNs in the
system we model. We implement Winograd on only AlexNet but we expect
Winograd to have limited benefits when used with other CNNs in our system.

We evaluate each CNN with two different use cases: 1) a real-time video
application that requires 30 frames per second (fps), and 2) a robotic
vision application that requires $60\,fps$. Thus, in our evaluation, the
accelerator in the system we model invokes CNN inference either at
$30\,fps$ or $60\,fps$, and we do not have any other performance
requirements. Because of this, although RTC can improve system performance
by eliminating a significant fraction of refresh operations and perform
more accesses instead, we do not quantitatively evaluate potential
performance benefits of RTC in the scope of this work.

\vspace{1mm}

\textbf{Tools, Technology, Area, and Energy Models.} We use commercial
EDA tools for all of our designs. We synthesize our designs to run
at $200\,MHz$ frequency using the $40\,nm$ technology node
for both CMOS and DRAM logic. To quantify memory controller area and
energy overhead of the RTC logic, we use a Micron-compatible
DRAM controller available from Gaisler~\cite{gaisler2013leon} as the
baseline. We extend this controller as we discuss in
Section~\ref{subsubsec:mctrl_changes}. We report area and energy
overheads based on post-layout data. The energy estimation for
the CMOS logic is based on gate-level simulation, back annotated with
post-layout data. To quantify area and energy overheads in DRAM, we
use the Rambus DRAM model~\cite{Vogelsang-micro2010} for different DRAM
dimensions and traces of access patterns. 

We create three different datapaths, one for each of the three
variants of RTC. For the full- and mid-RTC, we modify the
DRAM peripheral logic to reflect the RTC-enabled DRAM datapath.
For both models, we use technology parameters for $40\,nm$ DRAM from
ITRS~\cite{itrs-itrs2011}. By supplying the Rambus model a
trace of operations, in terms of activate, read, write, and
precharge, the Rambus model provides the energy numbers.
We generate traces using an in-house simulator~\cite{mocha}
for the workloads we evaluate. 

%\setstretch{0.978}
\vspace{-1mm}
\section{Evaluation}
\label{sec:results}

In this section, we analyze DRAM energy savings and area
overhead of each variant of RTC  compared to conventional low-power
DRAM, LPDDR4~\cite{jedec-lpddr4}. We evaluate six different
workloads in total. These include various CNNs (i.e., AlexNet (AN), LeNet
(LN), GoogleNet (GN), and ResNet50 (RN)), a Generative Adversarial Network
(GAN), and Winograd, which is an optimization for performing faster
convolution in CNNs. All these workloads vary in their memory
footprints and memory access patterns. We evaluate these
workloads on systems with different DRAM capacities. Furthermore, we
provide a breakdown of the benefits of the RTT  and PAAR techniques
that are part of the RTC framework.

\subsection{Energy Savings on Different Workloads}
\label{sec:energy_cnns}

We evaluate the DRAM energy savings of the three different implementations
of RTC on six CNN workloads in comparison to standard low-power
16\,Gb LPDDR4 DRAM. 

\begin{figure}[!ht] 
    \centering
    \begin{subfigure}[b]{.32\linewidth}
        \centering
        \includegraphics[width=\linewidth]{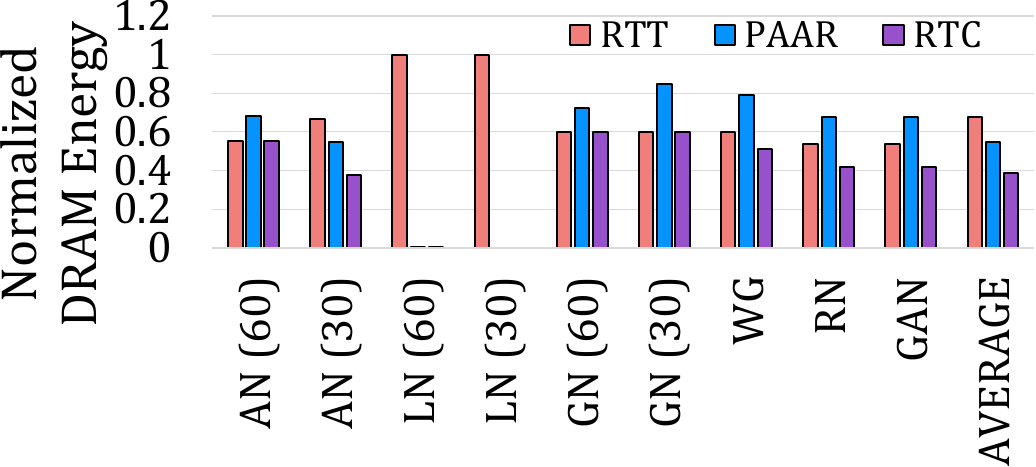}
        \caption{Full-RTC}
        \label{subfig:energy-fullrtc}
    \end{subfigure}
    \begin{subfigure}[b]{.32\linewidth}
        \centering
        \includegraphics[width=\linewidth]{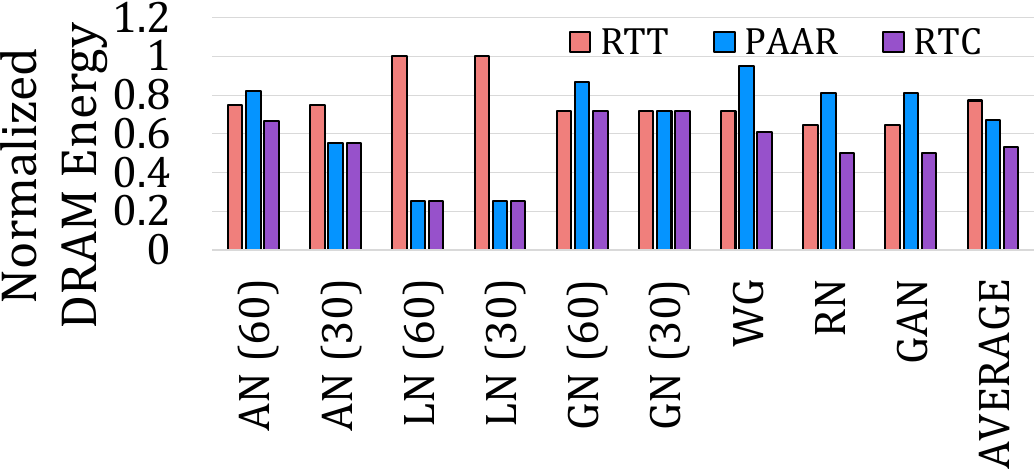}
        \caption{Mid-RTC}
        \label{subfig:energy-midrtc}
    \end{subfigure}
    \begin{subfigure}[b]{.32\linewidth}
        \centering
        \includegraphics[width=\linewidth]{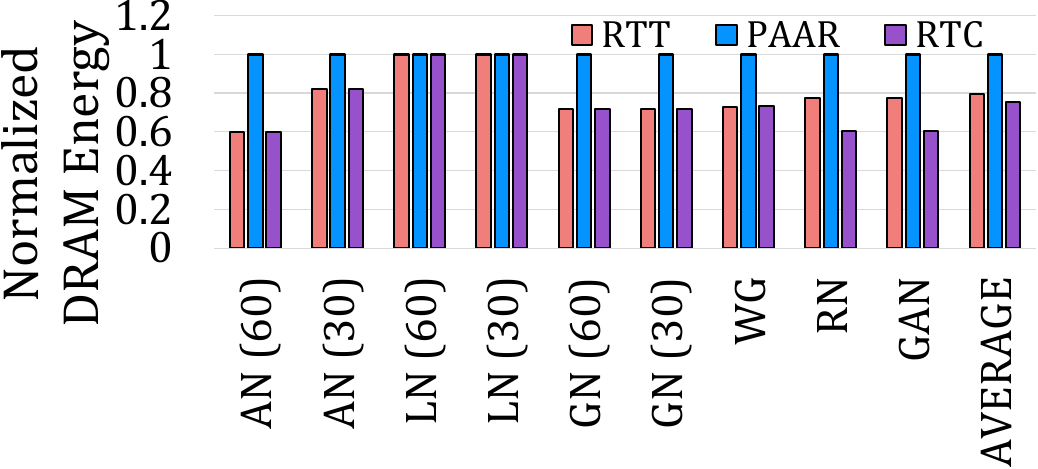}
        \caption{Min-RTC}
        \label{subfig:energy-minrtc}
    \end{subfigure}
    \caption{DRAM energy consumption of different RTC implementations
    normalized to the baseline LPDDR4 DRAM with standard refresh for
    AlexNet (AN), LeNet (LN), GoogleNet (GN), Winograd (WG), ResNet50 (RN),
    and Generative Adversarial Network (GAN)}. \label{fig:energy-rtc}
\end{figure}

\textbf{Full-RTC.} Figure~\ref{subfig:energy-fullrtc} plots DRAM energy
with full-RTC, normalized to the baseline DRAM with conventional refresh.
We break down the individual benefits of the RTT and PAAR techniques that
RTC combines. The DRAM energy savings of RTT primarily depend on how well
the DRAM refresh and access rates match: the closer they match, the greater
is the energy reduction. On average, RTT saves 32.3\% DRAM energy across
all workloads. RTT saves more DRAM energy at $60\,fps$ than at $30\,fps$
because running inference on the CNN more frequently results in a
larger number of DRAM accesses at $60\,fps$, which in turn creates
more opportunity for the accesses to align with the refreshes, and thus
makes the refreshes redundant. At $30\,fps$, the DRAM access rate is
almost the half of at $60\,fps$ and the refresh rate remains the same,
which results in an insufficient number of DRAM accesses to cover
all DRAM rows that contain workloads' data before the $64\,ms$ refresh
period. Therefore, at $30\,fps$, the memory controller needs to issue more
explicit refreshes that come with DRAM energy cost. Specifically, for
LeNet, the effectiveness of RTT is minimal because of the small memory
footprint and fewer read/write DRAM accesses of this workload. 

The PAAR technique saves DRAM energy by eliminating refreshes to DRAM
regions that are not allocated. Therefore, PAAR significantly favors
low-memory-footprint workloads, such as LeNet. PAAR alone saves 96\% DRAM
energy when running LeNet, as LeNet's working data set mostly fits
into accelerator's on-chip memory, and DRAM remains mostly idle. In such a
case, PAAR eliminates almost all refresh operations as very few DRAM rows
are allocated by LeNet. 

Full-RTC takes advantage of both RTT and PAAR at the same time and it
reduces DRAM energy consumption to 0.39x, achieving greater DRAM
energy savings than each technique achieves alone.

\textbf{Mid-RTC.} Figure~\ref{subfig:energy-midrtc} plots the DRAM energy
savings of mid-RTC. Mid-RTC implements a low-overhead version of PAAR that
operates at DRAM bank granularity, and thus PAAR in mid-RTC eliminates
refresh operations \emph{only} if a bank does not have any allocated rows.
As a result, mid-RTC PAAR saves less DRAM energy compared to full-RTC PAAR.
Similarly, mid-RTC implements a lighter version of RTT that is effective
\emph{only} when memory access rate is higher than refresh rate, i.e., when
memory accesses activate all rows that contain data at least once in every
$64\,ms$ refresh period. Mid-RTC reduces average DRAM energy
consumption to 0.53x compared to the baseline system. Therefore, mid-RTC
RTT is not as effective as full-RTC RTT, which can partially align memory
accesses with refreshes and issue explicit refresh only when
necessary.

\textbf{Min-RTC.} Figure~\ref{subfig:energy-minrtc} plots the DRAM energy
savings for min-RTC. Min-RTC is the most lightweight RTC implementation
that only employs the same RTT technique as in mid-RTC. On
average, it reduces DRAM energy consumption to 0.76x compared to the
baseline. Min-RTC provides the largest benefits for AlexNet at $60\,fps$,
reducing DRAM energy consumption by 40.0\%.

We conclude that all three variants of RTC save DRAM energy and the system
designer can choose the variant that fits best the energy and area
constraints.

\subsection{Sensitivity to DRAM Chip Capacity}
\label{sec:sensitivity_dram_density}

\begin{figure}[!ht] 
    \centering
    \includegraphics[width=.8\linewidth]{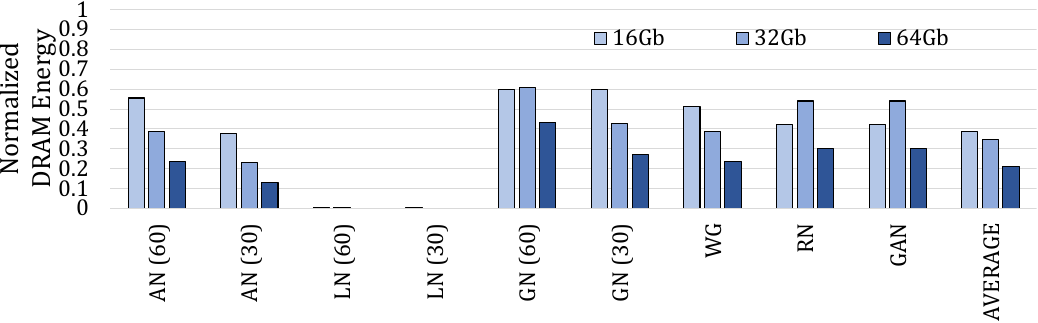}
    \caption{DRAM energy savings of Full-RTC when using DRAM chips with different densities.}
    \label{fig:dram_density_sensitivity}
\end{figure}

Figure~\ref{fig:dram_density_sensitivity} plots the energy savings of
full-RTC when employed in systems with different DRAM capacities. On
average, full-RTC provides higher DRAM energy saving as the DRAM capacity
increases, consuming 78.8\% less DRAM energy for 64\,Gb DRAM. This
is because high-capacity DRAM contains a large number of unallocated rows,
which PAAR skips refreshing. We note that two workloads, RN and GAN,
consume more energy with 32\,Gb DRAM than with 16\,Gb DRAM.
This is because the 32\,Gb DRAM we evaluate has the same number of
DRAM rows as the 16\,Gb DRAM but each row contains double the
number of DRAM cells compared to the 16\,Gb DRAM. We
notice that RN and GAN allocate slightly more DRAM rows
\emph{partially} when using 32\,Gb DRAM, which slightly increases
the DRAM energy consumption compared to 16\,Gb DRAM.

\subsection{Sensitivity to Data Locality Exploitation}
\label{sec:sensitivity_locality}

\emph{Data Locality exploitation} refers to the ability of the
system to cache the data read from DRAM. For example, a data locality
exploitation of 100\% implies that once the data is read from DRAM during
an iteration in a CNN layer, the data never leaves the CPU cache, and
thus it is not read from the DRAM again during the same iteration.
Similarly, a data locality exploitation of 50\% implies that
the data set is read twice from the DRAM during each iteration. For many
CNN applications, it is likely to achieve a data locality exploitation of
approximately 100\%, as reported in~\cite{chen2016eyeriss}.

\begin{figure}[!ht]
    \centering
    \includegraphics[width=.8\linewidth]{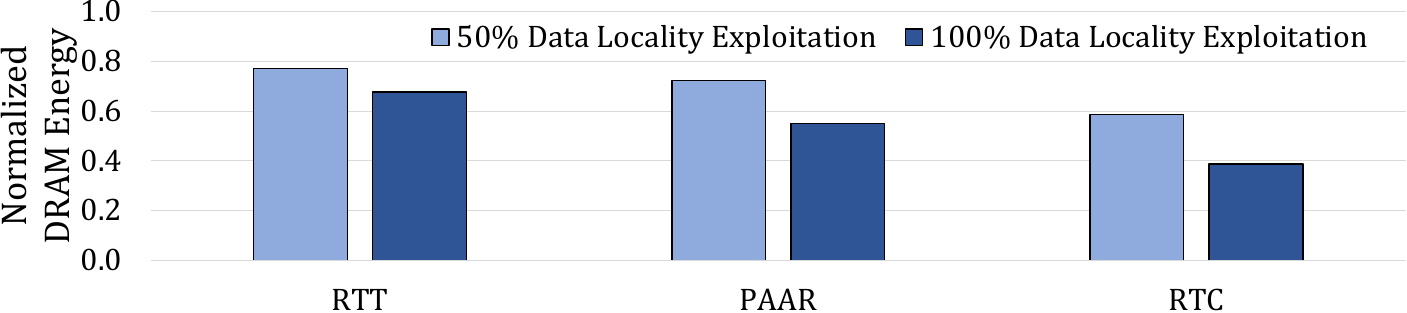}
    \caption{Average DRAM energy savings of full-RTC vs. data locality exploitation ratio.}
    \label{fig:locality_exploitation_sensitivity}
\end{figure}

We now elaborate on the impact of data locality exploitation on the
effectiveness of RTC. Figure~\ref{fig:locality_exploitation_sensitivity}
plots the normalized DRAM energy consumption for RTC with 50\% and 100\%
data locality exploitation. The absolute energy savings of the PAAR
components of RTC are not dependent on data locality exploitation. This is
because PAAR eliminates refreshes to unallocated regions in DRAM and the
rate at which allocated regions are accessed does not affect PAAR. However,
overall DRAM energy reduction with PAAR reduces when data locality
exploitation is low because frequent DRAM accesses increase the access
energy proportionally to the refresh energy, which remains constant.

The RTT component of RTC benefits more from low data locality
exploitation. As we explain in Section~\ref{subsec:simple_alignment}, RTT
eliminates refresh overhead when an application accesses its data
frequently enough. Therefore, low data locality exploitation causes the
DRAM to be accessed more frequently and this enables more of
the refresh requests to be synchronized with accesses. 

Overall, full-RTC saves 41.3\% and 61.3\% DRAM energy for 50\% and
100\% data locality exploitation, respectively.

\vspace{-2mm}
\subsection{Comparison to the Most Relevant Works} 
\begin{figure}[!ht]
\centering
    \includegraphics[width=0.50\linewidth]{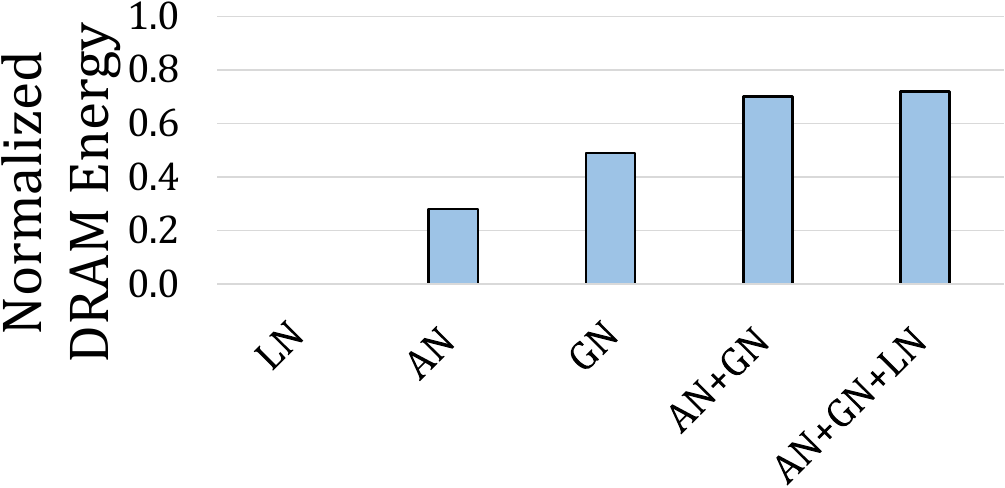}
    \vspace{-2mm}
    \caption{DRAM energy consumption with RTC normalized to DRAM
    energy consumption with SmartRefresh.}
    \label{fig:smartrefresh}
\end{figure}

In this section, we
compare our RTC mechanism against prior works that attempt to
reduce the DRAM refresh overhead. In particular, we compare our work with
SmartRefresh~\cite{ghosh2007smart}, the most closely-related work to RTC.
The key idea of SmartRefresh is to keep a history of the
recently-accessed rows and avoid refreshing these rows as their
cells' charge is already replenished when they were recently
accessed. SmartRefresh maintains 3-bit counters for each row.
Using the counters, it ensures that a row is not refreshed if it had been
accessed recently. To compare RTC against SmartRefresh, we implement a
DRAM controller with additional row counters (needed for
SmartRefresh).  For this evaluation, we assume an
$8\,GB$ DRAM module with a row size of $2048\,B$.  To utilize the DRAM
bandwidth, we run multiple instances of LeNet (LN), GoogleNet (GN),
and AlexNet (AN). We assume that each CNN requires operation at $60\,fps$.
We calculate the access patterns using state-of-the-art row
stationary data flow~\cite{chen2016eyeriss}.
Figure~\ref{fig:smartrefresh} shows the energy savings of RTC
over SmartRefresh. The figure shows that RTC provides from
28\% to 96\% energy reduction, compared to SmartRefresh.

RTC outperforms SmartRefresh using three optimizations to reduce the
refreshes. First, RTC aligns the refresh with reads. In this way it
ensures that the energy spent on both refresh and read is not wasted.
Second, RTC prevents the refresh of the DRAM rows that are \emph{not}
being used (i.e., not allocated). Third, RTC does not refresh the
rows that have been recently accessed. However, SmartRefresh applies
\emph{only} the third optimization by not refreshing recently-accessed
DRAM rows. As a result, SmartRefresh is ineffective when data transfer
rate is lower than the refresh rate, e.g., when only LeNet is running on
the system. In contrast, our RTC mechanism can reduce DRAM accesses
regardless of the data rate. SmartRefresh is effective when access rate is
greater than the refresh rate, which is the case in the rightmost two bar
graphs, where multiple workloads run together. However, even in these
two cases, RTC provides a significant $\approx 30\%$ DRAM energy reduction
over SmartRefresh. The main reason is the large number (e.g., 4,194,304 in
our evaluated system) of SRAM counters that SmartRefresh needs to maintain
to keep track of when each row is accessed. These counters consume a
significant amount of energy that offsets the benefits of refresh
reduction.

Refrint~\cite{agrawal2013refrint} is another refresh reduction
technique, which has the advantage of being effective for low
data access rates. However, Refrint has the downside of being applicable
to only embedded DRAM that is used as a cache. This is because Refrint is
based on the idea of refreshing only data that will be accessed in near
future and flushing the rest back to the main memory. In contrast, our
approach is generally applicable to any type of DRAM with small
changes in the DRAM chips.

Similar to our PAAR technique, ESKIMO~\cite{isen2009eskimo} skips
refreshes to unallocated memory regions. However, it does not perform any
refresh-access synchronization. Hence, ESKIMO does not reduce energy in
allocated regions of memory.

\subsection{Scalability Benefits}
Refresh is a growing major energy and performance
bottleneck with the scaling of the DRAM
technology~\cite{liu2012raidr, kang-memcon2014}. RTC mitigates this
negative scaling trend~\cite{liu2012raidr, venkatesan2006retention,
ghosh2007smart, chang2014improving, refreshnow, mukundan2013understanding,
reaper, mutlu2013memory, mutlu2015research, chou2015reducing,
khan2016parbor, khan2017detecting, khan-sigmetrics2014, patel2020bit,
patel2019understanding, kang-memcon2014, liu2013experimental} for a class
of applications by minimizing the need to refresh with its Refresh
Triggered Transfer (RTT) and Partial Array Auto Refresh (PAAR) techniques.
For a $64\,Gb$ DRAM chip, even when working at peak bandwidth,
refresh is expected to consume 46\% of the total DRAM energy
\cite{liu2012raidr,Jung-memsys2015}. To understand how RTC mitigates the
refresh overhead, consider two extremes of applications' DRAM access
characteristics. The first extreme is when the application has a small
data set. For this scenario, almost all the DRAM energy will be spent on
refresh. The PAAR technique eliminates this refresh overhead. It should be
noted that when the memory controller puts the DRAM into self-refresh
mode or power-down mode, PAAR still reduces DRAM energy
consumption since rows are still refreshed while conventional DRAM is
in one of these modes, and PAAR can eliminate unnecessary
refreshes. The second extreme is when the application utilizes the
\emph{entire} DRAM capacity and has a high DRAM bandwidth demand.
Note that, in this scenario, DRAM cannot switch to a low-power
mode as it needs to keep servicing access requests. In such a
scenario, conventional DRAM still spends a significant amount of
energy on refresh in addition to read/write accesses.
\cite{liu2012raidr,Jung-memsys2015} report that 47\% of the total DRAM
energy is spent while refreshing a DRAM chip of size
$64\,Gb$. However, in an RTC-enabled DRAM, a large
portion of refreshes can be eliminated by implicit refreshes
for applications that have regular memory access patterns.
Thus, in both extremes, RTC reduces the energy spent on
refresh, and thus provides better scalability of DRAM
in future technology nodes for applications with access patterns
that are amenable to it.  

\begin{figure}[!htbp]
    \centering
    \includegraphics[width=0.65\linewidth]{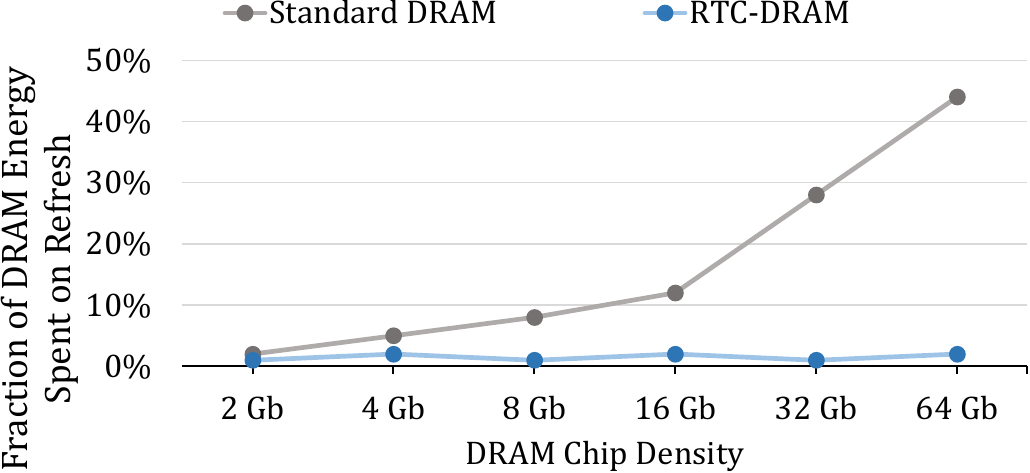}
    \caption{Fraction of DRAM energy spent on refresh as a function of DRAM chip capacity.}
    \vspace{-3mm}
    \label{fig:refreshoverhead}
\end{figure}

To make the above arguments more concrete, we quantify the scalability of
RTC for emerging large DRAMs when used for CNN applications. We perform an
experiment by utilizing the entire bandwidth of a DRAM module. We show our
results in Figure~\ref{fig:refreshoverhead}. It can be seen that
RTC-enabled DRAM almost completely eliminates the DRAM refresh energy
for CNN applications. Note that our results are consistent with prior
work~\cite{Jung-memsys2015,liu2012raidr}, providing external validity to
the experimental setup that we use. 

\subsection{Overhead of RTC}
RTC-enabled DRAM incurs almost none to modest area, energy, and latency
overheads. The area overhead mainly stems from 1) the
configurable refresh counter (see Section~\ref{sec:architecture}), 2)
the \emph{AGU for address generation}, 3) the modifications to the
data path (see Figure~\ref{fig:modifiedddr3}), 4) the RTC Backend
Controller (see Section~\ref{subsubsec:dram_changes}) and 5) the RTC
Frontend Controller. We quantify the area overhead by synthesizing
full-RTC in the standard CMOS $40\,nm$ technology. We synthesize the
corresponding logic components of the conventional DRAM also in the same
process since we do not have access to DRAM process technology
parameters. However, to approximate DRAM process in a more fair and
accurate way, we restrict the physical design tool to use only three
layers as commonly done in DRAM process. Our experiments show
that RTC has an area overhead of 0.18\% compared to a conventional $2\,Gb$
DRAM chip. This area overhead proportionally decreases as DRAM chip
density increases. This is because the area of only a few RTC
components (e.g., counters) increase with DRAM chip density, whereas the
area of a large number of components (e.g., RTC Backend Controller)
do not change.
 
The latency overhead of the RTC logic stems from the extra cycles needed to
reconfigure the RTC registers and state machines. However, the
latency overhead is negligibly small compared to the execution time
of a typical CNN-like application and reconfiguration of the RTC
logic likely occurs only once when an application starts.

\subsection{Using RTC with Non-CNN Workloads}
\label{sec:other_applications}

So far, we have focused on CNNs as an example for discussing and
quantifying the benefits and overhead of RTC. However, we believe RTC can
be applied to a wide variety of applications that have a regular
access pattern. We analyze the access patterns of three such well-known
applications and estimate the benefits of RTC while executing them. These
applications are: 1) Face recognition algorithm using
Eigenfaces~\cite{Kirby-TPAM1990}, 2) Bayesian Confidence Propagation Neural
Network (BCPNN), a spiking neural network model of biologically
plausible human brain cortex \cite{farahini-aspdac2014}, and 3) the
bioinformatics sequence alignment algorithm BFAST \cite{homer-plos2009}.
The reason for choosing these particular applications is that all of them
largely differ in DRAM access characteristics compared to CNNs. 

\begin{figure}[htbp]
    \centering
    \includegraphics[width=0.5\linewidth]{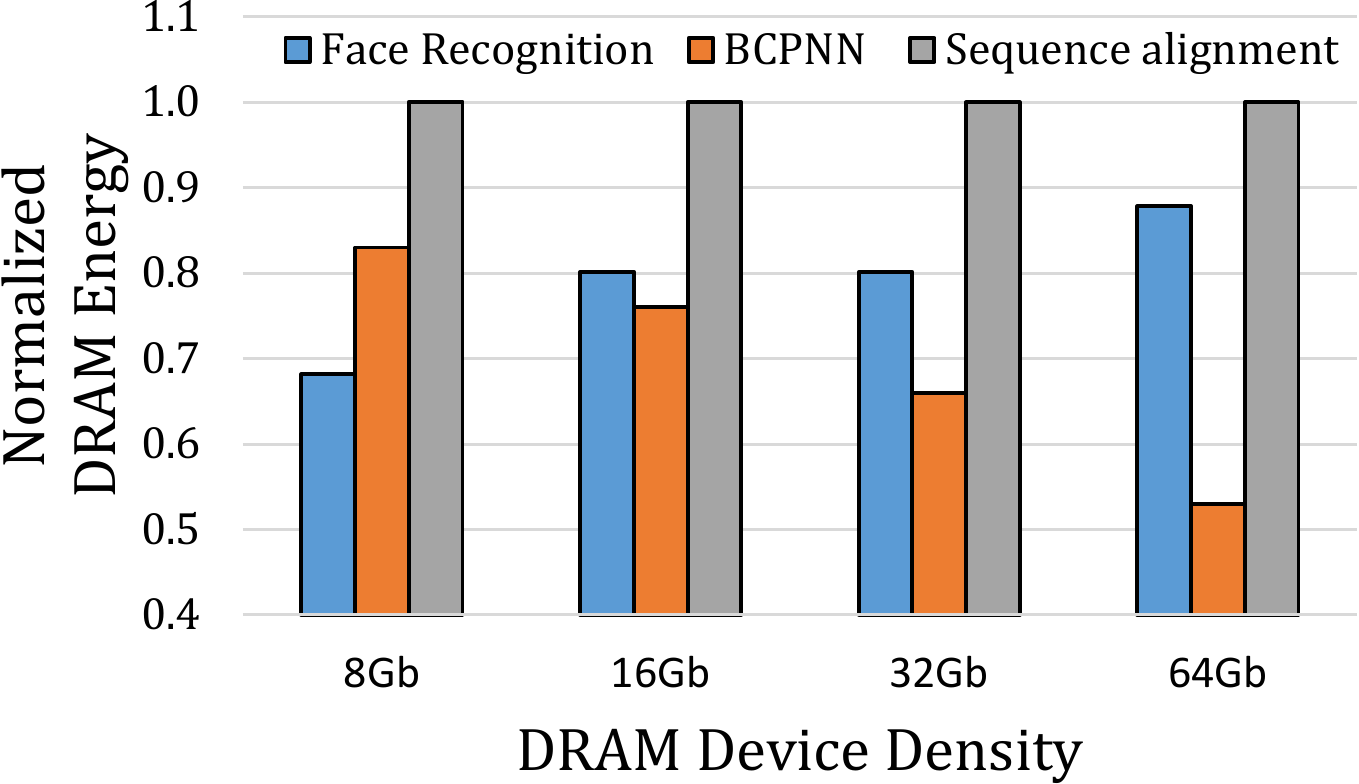}
    \caption{DRAM energy savings of RTC on applications from different domains}
    \label{fig:extentiontootherapplication}
\end{figure}

Figure~\ref{fig:extentiontootherapplication} shows the estimated DRAM
energy reduction for these three applications when using full-RTC-enabled
DRAM chips with different densities. Face recognition is a streaming
application that requires multiple filtering stages, which typically
access the same data multiple times from DRAM. We evaluate face
recognition using images of size $1024*1024*3$ and frame rate of $60\,fps$.
We find that full-RTC saves 12.2\% to 31.9\% DRAM energy for face
recognition depending on the DRAM chip density. 

BCPNN is a memory- and compute-intensive application that requires
approximately 740\,teraflop/s computational bandwidth and
30\,TB memory storage with a bandwidth of
112\,TB/s~\cite{farahini-aspdac2014}. During a single
iteration, the BCPNN workload accesses its entire allocated memory
four times. Because of such high rate of access to all of the
allocated memory, the RTT technique largely eliminates the need for refresh
in BCPNN, whereas PAAR provides small benefits since BCPNN allocates the
majority of the system's memory. Full-RTC saves 17.0\% to 47.3\% DRAM
energy for BCPNN depending on the DRAM chip density.

BFAST is based on the well-known Smith-Waterman local DNA sequence
alignment algorithm~\cite{smith1981identification}. BFAST has a mix
of random- and linear-access patterns. For this application, the RTC
circuitry is bypassed as neither PAAR nor RTT is effective.
By evaluating BFAST, we show that RTC can be disabled when
it is not effective, which incurs less than 0.01\% DRAM energy overhead, as
shown in the rightmost bars of
Figure~\ref{fig:extentiontootherapplication}.

\section{Related work}
\label{sec:relatedwork}

To our knowledge, this work is the first to methodically synchronize
applications' memory accesses with DRAM refreshes, so that the overhead
caused by refresh operations is significantly reduced in Convolutional
Neural Networks (CNNs). We briefly describe related work in DRAM refresh
optimization and CNN storage optimization.

\subsection{DRAM Refresh Optimization}

Several previous works change the DRAM refresh scheduling policy
to improve DRAM energy efficiency or performance. Bhati et
al.~\cite{bhatti-isca2015} present a flexible refresh mechanism to reduce
the refreshes.  Stuecheli et al.~\cite{stue-micro2010} propose a
technique that avoids interfering requests by altering the refresh
schedule. It delays a refresh depending on the number of postponed
refreshes and the predicted rank idle time.  Mukundan et
al.~\cite{mukundan2013understanding} propose various scheduling techniques
to tackle command queue contention. Chang et
al.~\cite{chang2014improving} provide mechanisms to parallelize accesses
and refreshes via scheduling and DRAM changes. However, all these
techniques consider refresh and memory access as two disjoint processes and
attempt to reduce the collisions between them as opposed to
synchronizing accesses and refreshes like we do. 

Various works~\cite{liu2012raidr, Liu:2011, venkatesan2006retention,
khan-sigmetrics2014, khan2016parbor, khan2016case, liu2013experimental,
qureshi-dsn2015, refreshnow, cui2014dtail, emma2008rethinking,
ghosh2007smart, khan2017detecting, reaper, isen2009eskimo,
luo2014characterizing, luo2020clr, kim2020charge, kim2020smart,
choi2020reducing, das2018vrl} reduce unnecessary refreshes by exploiting
the properties of DRAM cells and stored data. These works require
expensive mechanisms to discover the retention times of different DRAM
cells~\cite{liu2012raidr, qureshi-dsn2015, refreshnow, cui2014dtail,
emma2008rethinking, ghosh2007smart, khan-sigmetrics2014, khan2016parbor,
khan2016case, khan2017detecting, venkatesan2006retention, reaper,
liu2013experimental, luo2020clr, choi2020reducing, das2018vrl} or
require knowledge of how tolerant stored data is to retention
failures~\cite{cui2014dtail, isen2009eskimo, luo2014characterizing,
Liu:2011, kim2020charge, kim2020smart}. RTC does not require such
methods.

Zulian et al.~\cite{zulian2020access} propose a mechanism that
creates a mask of recently-accessed rows for each bank and introduces a
modified refresh command to skip refreshing the masked rows. Their
mechanism, likely concurrently developed with RTC, achieves a similar goal
as RTC but has a large area overhead as it stores one bit for every row in
DRAM.

SmartRefresh~\cite{ghosh2007smart}, Refrint~\cite{agrawal2013refrint}, and
Refree~\cite{pourshirazi2016refree} are techniques that reduce the refresh
overhead based on the memory access patterns of applications. These
techniques are closely related to RTC.  SmartRefresh~\cite{ghosh2007smart}
reduces refresh energy in DRAM by maintaining a timeout counter
for each row. This mechanism avoids unnecessary refreshes of recently
accessed rows.  However, SmartRefresh does not skip refreshing rows
that do not store useful data. Thus, SmartRefresh is not
effective for applications that have a small memory footprint where a
significant number of DRAM rows do not contain useful data. Furthermore,
SmartRefresh requires significant additional energy to maintain
the large number of counters (see Section~\ref{sec:results}).
$Refrint$~\cite{agrawal2013refrint} eliminates refresh to unused DRAM
rows. However, its overheads are evaluated only for embedded DRAMs.
Implementing this technique on off-chip DRAMs would require changing
the memory arrays (i.e., it would be even more invasive than
Full-RTC). Furthermore, similar to SmartRefresh, Refrint suffers from
the overhead of maintaining the state of each DRAM row.
Refree~\cite{pourshirazi2016refree} combines a non-volatile PCM
memory with conventional DRAM to eliminate DRAM refresh by moving a
row to PCM when the row needs to be refreshed. Refree requires
retention timeout counters and incurs overhead of moving data between PCM
and DRAM. Compared to these approaches, RTC does not require any per row
state. RTC improves the energy efficiency with small overhead on the
DRAM chip and the memory controller.

ESKIMO~\cite{isen2009eskimo} eliminates refreshes in unallocated
memory regions. However, ESKIMO does not synchronize memory accesses with
refreshes, and thus it does not reduce refresh energy in memory regions
that allocate data. Using RTT and PAAR, RTC reduces the energy overhead of
refresh operations on both allocated and unallocated portions of the
memory.

\subsection{CNN Storage Optimization}

Driven by the success of CNNs as a machine learning technique, many
researchers have focused on implementation aspects of CNN. While
initially researchers focused on speeding up and improving the energy
efficiency of the computational
aspects~\cite{Qadeer-ISCA2013,Temam-ISCA2012,Esmaeil-MICRO2012}, recently,
the research have shifted towards improving the efficiency of the
memory~\cite{chen2014diannao,chen2016eyeriss,dally2016eie}.

Chen et al.~\cite{chen2014diannao} show that CNNs can be viewed as nested
loops. They present an accelerator that reduces memory footprint using loop
tiling. Du et al.~\cite{du2015shidiannao} build on top
of~\cite{chen2014diannao} and propose an accelerator architecture that uses
only SRAM to store application data, eliminating DRAM completely. While
their approach is applicable some application domains, many
accelerators~\cite{dally2016eie,chen2016eyeriss} are designed to work with
a DRAM to meet the memory requirements of large neural networks. Chen et
al.~\cite{chen2016eyeriss} show a technique to optimize the data movement
between the memory and the computational units. Song et
al.~\cite{dally2016eie} present a technique to reduce the number of memory
accesses using compression in classification layers. However, even after
fully exploiting data locality, most of the energy is still spent on data
transfers between DRAM and SRAM. Overall, these prior works aim to
mitigate DRAM overhead in NN applications by exploiting data locality to
better utilize SRAM-based memories. However, such techniques do not reduce
DRAM refresh energy, and thus, DRAM refresh incurs significant overhead.

RANA~\cite{tu2018rana} employs embedded DRAM (eDRAM) as an additional
on-chip buffer to SRAM. RANA mitigates the refresh overhead of eDRAM by
disabling refresh when data lifetime in an eDRAM bank is shorter
than the retention time of the DRAM. RTC is complementary to this
work as RTC mitigates the refresh overhead when data stored in DRAM has a
long lifetime by synchronizing accesses to data with refresh operations. 

EDEN~\cite{koppula2019eden} implements energy-efficient approximate DRAM
for neural network inference by exploiting the error tolerance property of
neural networks. EDEN has the limitation of being only applicable
to data that has error tolerance. In contrast, RTC can mitigate DRAM
refresh without causing bit flips due to retention failures in DRAM.
EDEN and RTC can be combined for higher energy savings than each can
achieve alone.

To the best of our knowledge, RTC is the first work that provides
architectural solution for mitigating DRAM refresh energy in CNNs by
synchronizing applications' memory accesses with DRAM refresh operations.

\section{Future Work}
\label{sec:future_work}

We envision at least two major avenues of future work. 

First, we plan to evaluate more applications and check if they can benefit
from the new Refresh Triggered Computation model we propose. For
example, we expect various applications from domains such as deep learning,
computer vision, bioinformatics, and high-performance computing to highly
benefit from RTC. We believe RTC has the potential to be applied to a wide
range of applications from a variety of domains, as long as the
access patterns can be regularized and synchronized with refresh.

Second, we plan to use RTC with a new class of neural networks,
\emph{Self-Organizing Maps} that a prior work~\cite{yang2018ribosom} uses
for rapid and accurate identification of bacterial genomes and their
resistance to antibiotics. We also plan to incorporate the RTC technique as
a part of the SiLago~\cite{hemani2017synchoricity} framework, which is a
Lego-inspired VLSI design framework that we develop. We plan to expand such
a synthesis framework to map multiple complex workloads to custom SiLago
design instances that will use DRAM enhanced with RTC as main memory.

\section{Conclusion}
\label{sec:conclusion}

We introduce a new software/hardware cooperative DRAM
refresh optimization technique, which we refer to as Refresh Triggered
Computation (RTC).  RTC significantly reduces DRAM refresh overhead
using two key concepts. First, it synchronizes DRAM refreshes
with application read/write accesses to reduce the number of required
refresh operations by exploiting the fact that application DRAM
accesses implicitly replenish the charge of the DRAM cells. Second,
RTC eliminates refreshing of rows that do not have any data
allocated. We propose three variants of RTC, which differ in the
level of area overhead incurred in the memory controller and the DRAM
chip. Our extensive evaluations using commonly-used
Convolutional Neural Networks (CNNs) show that the most aggressive
variant of RTC reduces average DRAM energy by 61.3\% while
incurring only 0.18\% area overhead over a conventional DRAM chip. We
also show that RTC improves DRAM energy consumption of workloads from
different domains. We conclude that RTC largely mitigates DRAM
refresh overhead in both CNN applications and various other
applications by synchronizing applications' DRAM accesses with DRAM
refresh operations. We hope that RTC inspires other software/hardware
cooperative mechanisms to reduce DRAM energy in data-intensive workloads.

\section*{Acknowledgments}

We thank the anonymous TACO 2020 reviewers for their feedback and the
SAFARI group members for the stimulating intellectual environment they
provide. We acknowledge the generous gifts provided by our industrial
partners: ASML, Facebook, Google, Huawei, Intel, Microsoft, and VMware.
This research was supported in part by the Semiconductor Research
Corporation. An earlier version of this article was placed on
arxiv.org~\cite{jafri2019refresh} in October 2019.

%\setstretch{.97}
 
%%%%%%% -- PAPER CONTENT ENDS -- %%%%%%%%

%%%%%%%%% -- BIB STYLE AND FILE -- %%%%%%%%
\bibliographystyle{ACM-Reference-Format-num}
\bibliography{refs}


\begin{thebibliography}{121}

%%% the Web address.

\ifx \showCODEN    \undefined \def \showCODEN     #1{\unskip}     \fi
\ifx \showDOI      \undefined \def \showDOI       #1{#1}\fi
\ifx \showISBNx    \undefined \def \showISBNx     #1{\unskip}     \fi
\ifx \showISBNxiii \undefined \def \showISBNxiii  #1{\unskip}     \fi
\ifx \showISSN     \undefined \def \showISSN      #1{\unskip}     \fi
\ifx \showLCCN     \undefined \def \showLCCN      #1{\unskip}     \fi
\ifx \shownote     \undefined \def \shownote      #1{#1}          \fi
\ifx \showarticletitle \undefined \def \showarticletitle #1{#1}   \fi
\ifx \showURL      \undefined \def \showURL       {\relax}        \fi
% invisible to TeX
\providecommand\bibfield[2]{#2}
\providecommand\bibinfo[2]{#2}
\providecommand\natexlab[1]{#1}
\providecommand\showeprint[2][]{arXiv:#2}

\bibitem[\protect\citeauthoryear{Agrawal, Jain, Ansari, and Torrellas}{Agrawal
  et~al\mbox{.}}{2013}]%
        {agrawal2013refrint}
\bibfield{author}{\bibinfo{person}{A.~Agrawal}, {et~al\mbox{.}}}
  \bibinfo{year}{2013}\natexlab{}.
\newblock \showarticletitle{{Refrint: Intelligent Refresh to Minimize Power in
  on-chip Multiprocessor Cache Hierarchies}}. In
  \bibinfo{booktitle}{\emph{HPCA}}.
\newblock


\bibitem[\protect\citeauthoryear{{AMBA Specification}}{{AMBA
  Specification}}{1999}]%
        {specification1999rev}
\bibfield{author}{\bibinfo{person}{{AMBA Specification}}.}
  \bibinfo{year}{1999}\natexlab{}.
\newblock \showarticletitle{Rev. 2.0}.
\newblock \bibinfo{journal}{\emph{ARM, http://www.arm.com}}
  (\bibinfo{year}{1999}).
\newblock


\bibitem[\protect\citeauthoryear{Baek, Cho, and Melhem}{Baek
  et~al\mbox{.}}{2014}]%
        {refreshnow}
\bibfield{author}{\bibinfo{person}{S.~Baek}, {et~al\mbox{.}}}
  \bibinfo{year}{2014}\natexlab{}.
\newblock \showarticletitle{{Refresh Now and Then}}.
\newblock \bibinfo{journal}{\emph{IEEE TC}} (\bibinfo{year}{2014}).
\newblock


\bibitem[\protect\citeauthoryear{Bhati, Chishti, Lu, and Jacob}{Bhati
  et~al\mbox{.}}{2015}]%
        {bhatti-isca2015}
\bibfield{author}{\bibinfo{person}{I.~Bhati}, {et~al\mbox{.}}}
  \bibinfo{year}{2015}\natexlab{}.
\newblock \showarticletitle{{Flexible Auto-Refresh: Enabling Scalable and
  Energy-Efficient DRAM Refresh Reductions}}. In
  \bibinfo{booktitle}{\emph{ISCA}}.
\newblock


\bibitem[\protect\citeauthoryear{Chabloz and Hemani}{Chabloz and
  Hemani}{2014}]%
        {jeanvtvlsi2014}
\bibfield{author}{\bibinfo{person}{J.~M. Chabloz} {et~al\mbox{.}}}
  \bibinfo{year}{2014}\natexlab{}.
\newblock \showarticletitle{{Low-Latency Maximal-Throughput Communication
  Interfaces for Rationally Related Clock Domains}}.
\newblock \bibinfo{journal}{\emph{TVLSI}} (\bibinfo{year}{2014}).
\newblock


\bibitem[\protect\citeauthoryear{Chang, Kashyap, Hassan, Ghose, Hsieh,
  et~al\mbox{.}}{Chang et~al\mbox{.}}{2016a}]%
        {chang2016understanding}
\bibfield{author}{\bibinfo{person}{K.~K. Chang}, {et~al\mbox{.}}}
  \bibinfo{year}{2016}\natexlab{a}.
\newblock \showarticletitle{{Understanding Latency Variation in Modern {DRAM}
  Chips: Experimental Characterization, Analysis, and Optimization}}. In
  \bibinfo{booktitle}{\emph{SIGMETRICS}}.
\newblock


\bibitem[\protect\citeauthoryear{Chang, Nair, Lee, Ghose, Qureshi,
  et~al\mbox{.}}{Chang et~al\mbox{.}}{2016b}]%
        {chang2016low}
\bibfield{author}{\bibinfo{person}{K.~K. Chang}, {et~al\mbox{.}}}
  \bibinfo{year}{2016}\natexlab{b}.
\newblock \showarticletitle{{Low-Cost Inter-Linked Subarrays (LISA): Enabling
  Fast Inter-Subarray Data Movement in DRAM}}. In
  \bibinfo{booktitle}{\emph{HPCA}}.
\newblock


\bibitem[\protect\citeauthoryear{Chang, Ya{\u{g}}l{\i}k{\c{c}}{\i}, Ghose,
  Agrawal, Chatterjee, et~al\mbox{.}}{Chang et~al\mbox{.}}{2017}]%
        {chang2017understanding}
\bibfield{author}{\bibinfo{person}{K.~K. Chang}, {et~al\mbox{.}}}
  \bibinfo{year}{2017}\natexlab{}.
\newblock \showarticletitle{{Understanding Reduced-Voltage Operation in Modern
  DRAM Devices: Experimental Characterization, Analysis, and Mechanisms}}. In
  \bibinfo{booktitle}{\emph{SIGMETRICS}}.
\newblock


\bibitem[\protect\citeauthoryear{Chang, Lee, Chishti, Alameldeen, Wilkerson,
  et~al\mbox{.}}{Chang et~al\mbox{.}}{2014}]%
        {chang2014improving}
\bibfield{author}{\bibinfo{person}{K.~K.-w. Chang}, {et~al\mbox{.}}}
  \bibinfo{year}{2014}\natexlab{}.
\newblock \showarticletitle{{Improving {DRAM} Performance by Parallelizing
  Refreshes with Accesses}}. In \bibinfo{booktitle}{\emph{HPCA}}.
\newblock


\bibitem[\protect\citeauthoryear{Chen, Du, Sun, Wang, Wu, et~al\mbox{.}}{Chen
  et~al\mbox{.}}{2014}]%
        {chen2014diannao}
\bibfield{author}{\bibinfo{person}{T.~Chen}, {et~al\mbox{.}}}
  \bibinfo{year}{2014}\natexlab{}.
\newblock \showarticletitle{{{DianNao}: A Small-footprint High-throughput
  Accelerator for Ubiquitous Machine-learning}}. In
  \bibinfo{booktitle}{\emph{ASPLOS}}.
\newblock


\bibitem[\protect\citeauthoryear{Chen, Emer, and Sze}{Chen
  et~al\mbox{.}}{2016}]%
        {chen2016eyeriss}
\bibfield{author}{\bibinfo{person}{Y.~Chen}, {et~al\mbox{.}}}
  \bibinfo{year}{2016}\natexlab{}.
\newblock \showarticletitle{{Eyeriss: A Spatial Architecture for
  Energy-Efficient Dataflow for Convolutional Neural Networks}}. In
  \bibinfo{booktitle}{\emph{ISCA}}.
\newblock


\bibitem[\protect\citeauthoryear{Choi, Hong, Lee, and Yoo}{Choi
  et~al\mbox{.}}{2020}]%
        {choi2020reducing}
\bibfield{author}{\bibinfo{person}{H.~Choi}, {et~al\mbox{.}}}
  \bibinfo{year}{2020}\natexlab{}.
\newblock \showarticletitle{{Reducing DRAM Refresh Power Consumption by Runtime
  Profiling of Retention Time and Dual-Row Activation}}.
\newblock \bibinfo{journal}{\emph{Microprocessors and Microsystems}}
  (\bibinfo{year}{2020}).
\newblock


\bibitem[\protect\citeauthoryear{Chou, Nair, and Qureshi}{Chou
  et~al\mbox{.}}{2015}]%
        {chou2015reducing}
\bibfield{author}{\bibinfo{person}{C.~Chou}, {et~al\mbox{.}}}
  \bibinfo{year}{2015}\natexlab{}.
\newblock \showarticletitle{{Reducing Refresh Power in Mobile Devices with
  Morphable ECC}}. In \bibinfo{booktitle}{\emph{DSN}}.
\newblock


\bibitem[\protect\citeauthoryear{{Cobham}}{{Cobham}}{2017}]%
        {gaisler2013leon}
\bibfield{author}{\bibinfo{person}{{Cobham}}.} \bibinfo{year}{2017}\natexlab{}.
\newblock \bibinfo{title}{{GRLIB IP Library User's Manual}}.
\newblock
  \bibinfo{howpublished}{http://www.gaisler.com/products/grlib/grlib.pdf}.
\newblock


\bibitem[\protect\citeauthoryear{Cui, McKee, Zha, Bao, and Chen}{Cui
  et~al\mbox{.}}{2014}]%
        {cui2014dtail}
\bibfield{author}{\bibinfo{person}{Z.~Cui}, {et~al\mbox{.}}}
  \bibinfo{year}{2014}\natexlab{}.
\newblock \showarticletitle{{DTail: A Flexible Approach to DRAM Refresh
  Management}}. In \bibinfo{booktitle}{\emph{ICS}}.
\newblock


\bibitem[\protect\citeauthoryear{Das, Hassan, and Mutlu}{Das
  et~al\mbox{.}}{2018}]%
        {das2018vrl}
\bibfield{author}{\bibinfo{person}{A.~Das}, {et~al\mbox{.}}}
  \bibinfo{year}{2018}\natexlab{}.
\newblock \showarticletitle{{VRL-DRAM: Improving DRAM Performance via Variable
  Refresh Latency.}}. In \bibinfo{booktitle}{\emph{DAC}}.
\newblock


\bibitem[\protect\citeauthoryear{Du, Fasthuber, Chen, Ienne, Li,
  et~al\mbox{.}}{Du et~al\mbox{.}}{2015}]%
        {du2015shidiannao}
\bibfield{author}{\bibinfo{person}{Z.~Du}, {et~al\mbox{.}}}
  \bibinfo{year}{2015}\natexlab{}.
\newblock \showarticletitle{{ShiDianNao: Shifting Vision Processing Closer to
  the Sensor}}. In \bibinfo{booktitle}{\emph{ISCA}}.
\newblock


\bibitem[\protect\citeauthoryear{Emma, Reohr, and Meterelliyoz}{Emma
  et~al\mbox{.}}{2008}]%
        {emma2008rethinking}
\bibfield{author}{\bibinfo{person}{P.~G. Emma}, {et~al\mbox{.}}}
  \bibinfo{year}{2008}\natexlab{}.
\newblock \showarticletitle{{Rethinking Refresh: Increasing Availability and
  Reducing Power in DRAM for Cache Applications}}.
\newblock \bibinfo{journal}{\emph{IEEE Micro}} (\bibinfo{year}{2008}).
\newblock


\bibitem[\protect\citeauthoryear{Esmaeilzadeh, Sampson, Ceze, and
  Burger}{Esmaeilzadeh et~al\mbox{.}}{2012}]%
        {Esmaeil-MICRO2012}
\bibfield{author}{\bibinfo{person}{H.~Esmaeilzadeh}, {et~al\mbox{.}}}
  \bibinfo{year}{2012}\natexlab{}.
\newblock \showarticletitle{{Neural Acceleration for General-Purpose
  Approximate Programs}}. In \bibinfo{booktitle}{\emph{MICRO}}.
\newblock


\bibitem[\protect\citeauthoryear{Farahini, Hemani, Lansner, Clermidy, and
  Svensson}{Farahini et~al\mbox{.}}{2014a}]%
        {farahini-aspdac2014}
\bibfield{author}{\bibinfo{person}{N.~Farahini}, {et~al\mbox{.}}}
  \bibinfo{year}{2014}\natexlab{a}.
\newblock \showarticletitle{{A Scalable Custom Simulation Machine for the
  Bayesian Confidence Propagation Neural Network Model of the Brain}}. In
  \bibinfo{booktitle}{\emph{ASP-DAC}}.
\newblock


\bibitem[\protect\citeauthoryear{Farahini, Hemani, Sohofi, Jafri, Tajammul,
  et~al\mbox{.}}{Farahini et~al\mbox{.}}{2014b}]%
        {farahini2014parallel}
\bibfield{author}{\bibinfo{person}{N.~Farahini}, {et~al\mbox{.}}}
  \bibinfo{year}{2014}\natexlab{b}.
\newblock \showarticletitle{{Parallel Distributed Scalable Runtime Address
  Generation Scheme for a Coarse Grain Reconfigurable Computation and Storage
  Fabric}}.
\newblock \bibinfo{journal}{\emph{Microprocessors and Microsystems}}
  (\bibinfo{year}{2014}).
\newblock


\bibitem[\protect\citeauthoryear{Ghose, Li, Hajinazar, Cali, and Mutlu}{Ghose
  et~al\mbox{.}}{2019}]%
        {ghose2019demystifying}
\bibfield{author}{\bibinfo{person}{S.~Ghose}, {et~al\mbox{.}}}
  \bibinfo{year}{2019}\natexlab{}.
\newblock \showarticletitle{{Demystifying Complex Workload-DRAM Interactions:
  An Experimental Study}}. In \bibinfo{booktitle}{\emph{SIGMETRICS}}.
\newblock


\bibitem[\protect\citeauthoryear{Ghose, Ya{\u{g}}l{\i}k{\c{c}}{\i}, Gupta, Lee,
  Kudrolli, et~al\mbox{.}}{Ghose et~al\mbox{.}}{2018}]%
        {ghose2018your}
\bibfield{author}{\bibinfo{person}{S.~Ghose}, {et~al\mbox{.}}}
  \bibinfo{year}{2018}\natexlab{}.
\newblock \showarticletitle{{What Your DRAM Power Models Are Not Telling You:
  Lessons from a Detailed Experimental Study}}. In
  \bibinfo{booktitle}{\emph{SIGMETRICS}}.
\newblock


\bibitem[\protect\citeauthoryear{Ghosh and Lee}{Ghosh and Lee}{2007}]%
        {ghosh2007smart}
\bibfield{author}{\bibinfo{person}{M.~Ghosh} {et~al\mbox{.}}}
  \bibinfo{year}{2007}\natexlab{}.
\newblock \showarticletitle{{Smart Refresh: An Enhanced Memory Controller
  Design for Reducing Energy in Conventional and 3D Die-stacked DRAMs}}. In
  \bibinfo{booktitle}{\emph{MICRO}}.
\newblock


\bibitem[\protect\citeauthoryear{Goodfellow, Pouget-Abadie, Mirza, Xu,
  Warde-Farley, et~al\mbox{.}}{Goodfellow et~al\mbox{.}}{2014}]%
        {goodfellow2014generative}
\bibfield{author}{\bibinfo{person}{I.~Goodfellow}, {et~al\mbox{.}}}
  \bibinfo{year}{2014}\natexlab{}.
\newblock \showarticletitle{{Generative Adversarial Nets}}. In
  \bibinfo{booktitle}{\emph{NIPS}}.
\newblock


\bibitem[\protect\citeauthoryear{Hassan, Patel, Kim,
  Ya{\u{g}}l{\i}k{\c{c}}{\i}, Vijaykumar, et~al\mbox{.}}{Hassan
  et~al\mbox{.}}{2019}]%
        {hassan2019crow}
\bibfield{author}{\bibinfo{person}{H.~Hassan}, {et~al\mbox{.}}}
  \bibinfo{year}{2019}\natexlab{}.
\newblock \showarticletitle{{CROW: A Low-Cost Substrate for Improving DRAM
  Performance, Energy Efficiency, and Reliability}}. In
  \bibinfo{booktitle}{\emph{ISCA}}.
\newblock


\bibitem[\protect\citeauthoryear{Hassan, Pekhimenko, Vijaykumar, Seshadri, Lee,
  et~al\mbox{.}}{Hassan et~al\mbox{.}}{2016}]%
        {hassan2016chargecache}
\bibfield{author}{\bibinfo{person}{H.~Hassan}, {et~al\mbox{.}}}
  \bibinfo{year}{2016}\natexlab{}.
\newblock \showarticletitle{{ChargeCache: Reducing DRAM Latency by Exploiting
  Row Access Locality}}. In \bibinfo{booktitle}{\emph{HPCA}}.
\newblock


\bibitem[\protect\citeauthoryear{Hassan, Vijaykumar, Khan, Ghose, Chang,
  et~al\mbox{.}}{Hassan et~al\mbox{.}}{2017}]%
        {hassan2017softmc}
\bibfield{author}{\bibinfo{person}{H.~Hassan}, {et~al\mbox{.}}}
  \bibinfo{year}{2017}\natexlab{}.
\newblock \showarticletitle{{SoftMC: A Flexible and Practical Open-Source
  Infrastructure for Enabling Experimental DRAM Studies}}. In
  \bibinfo{booktitle}{\emph{HPCA}}.
\newblock


\bibitem[\protect\citeauthoryear{He, Zhang, Ren, and Sun}{He
  et~al\mbox{.}}{2016}]%
        {he2016deep}
\bibfield{author}{\bibinfo{person}{K.~He}, {et~al\mbox{.}}}
  \bibinfo{year}{2016}\natexlab{}.
\newblock \showarticletitle{{Deep Residual Learning for Image Recognition}}. In
  \bibinfo{booktitle}{\emph{CVPR}}.
\newblock


\bibitem[\protect\citeauthoryear{Hemani, Jafri, and Masoumian}{Hemani
  et~al\mbox{.}}{2017}]%
        {hemani2017synchoricity}
\bibfield{author}{\bibinfo{person}{A.~Hemani}, {et~al\mbox{.}}}
  \bibinfo{year}{2017}\natexlab{}.
\newblock \showarticletitle{{Synchoricity and NOCs Could Make Billion Gate
  Custom Hardware Centric SOCs Affordable}}. In
  \bibinfo{booktitle}{\emph{NOCS}}.
\newblock


\bibitem[\protect\citeauthoryear{Homer, Merriman, and Nelson}{Homer
  et~al\mbox{.}}{2009}]%
        {homer-plos2009}
\bibfield{author}{\bibinfo{person}{N.~Homer}, {et~al\mbox{.}}}
  \bibinfo{year}{2009}\natexlab{}.
\newblock \showarticletitle{{BFAST: An Alignment Tool for Large Scale Genome
  Resequencing}}.
\newblock \bibinfo{journal}{\emph{PLoS ONE}} (\bibinfo{year}{2009}).
\newblock


\bibitem[\protect\citeauthoryear{Ili{\'c} and Stoj{\v{c}}ev}{Ili{\'c} and
  Stoj{\v{c}}ev}{2011}]%
        {ilic2011address}
\bibfield{author}{\bibinfo{person}{M.~Ili{\'c}} {et~al\mbox{.}}}
  \bibinfo{year}{2011}\natexlab{}.
\newblock \showarticletitle{{Address Generation Unit as Accelerator Block in
  DSP}}. In \bibinfo{booktitle}{\emph{TELSIKS}}.
\newblock


\bibitem[\protect\citeauthoryear{Instruments}{Instruments}{2002}]%
        {instruments2002tms320c55x}
\bibfield{author}{\bibinfo{person}{T.~Instruments}.}
  \bibinfo{year}{2002}\natexlab{}.
\newblock \showarticletitle{{TMS320C55x DSP Mnemonic Instruction Set Reference
  Guide}}.
\newblock \bibinfo{journal}{\emph{Literature Number: SPRU374G October}}
  (\bibinfo{year}{2002}).
\newblock


\bibitem[\protect\citeauthoryear{Ipek, Mutlu, Mart{\'\i}nez, and Caruana}{Ipek
  et~al\mbox{.}}{2008}]%
        {ipek2008self}
\bibfield{author}{\bibinfo{person}{E.~Ipek}, {et~al\mbox{.}}}
  \bibinfo{year}{2008}\natexlab{}.
\newblock \showarticletitle{{Self-Optimizing Memory Controllers: A
  Reinforcement Learning Approach}}. In \bibinfo{booktitle}{\emph{ISCA}}.
\newblock


\bibitem[\protect\citeauthoryear{Isen and John}{Isen and John}{2009}]%
        {isen2009eskimo}
\bibfield{author}{\bibinfo{person}{C.~Isen} {et~al\mbox{.}}}
  \bibinfo{year}{2009}\natexlab{}.
\newblock \showarticletitle{{ESKIMO - Energy Savings Using Semantic Knowledge
  of Inconsequential Memory Occupancy for DRAM Subsystem}}. In
  \bibinfo{booktitle}{\emph{MICRO}}.
\newblock


\bibitem[\protect\citeauthoryear{Itoh}{Itoh}{2013}]%
        {itoh2013vlsi}
\bibfield{author}{\bibinfo{person}{K.~Itoh}.} \bibinfo{year}{2013}\natexlab{}.
\newblock \bibinfo{booktitle}{\emph{{VLSI Memory Chip Design}}}.
  Vol.~\bibinfo{volume}{5}.
\newblock \bibinfo{publisher}{Springer Science \& Business Media}.
\newblock


\bibitem[\protect\citeauthoryear{ITRS}{ITRS}{2011}]%
        {itrs-itrs2011}
\bibfield{author}{\bibinfo{person}{ITRS}.} \bibinfo{year}{2011}\natexlab{}.
\newblock \bibinfo{title}{{International Technology Roadmap for Semiconductors
  2011 Edition: Executive Summary}}.
\newblock
  \bibinfo{howpublished}{http://www.itrs.net/Links/2011ITRS/2011Chapters/2011ExecSum.pdf}.
\newblock


\bibitem[\protect\citeauthoryear{Jacob, Ng, and Wang}{Jacob
  et~al\mbox{.}}{2010}]%
        {jacob2010memory}
\bibfield{author}{\bibinfo{person}{B.~Jacob}, {et~al\mbox{.}}}
  \bibinfo{year}{2010}\natexlab{}.
\newblock \bibinfo{booktitle}{\emph{{Memory Systems: Cache, DRAM, Disk}}}.
\newblock \bibinfo{publisher}{Morgan Kaufmann}.
\newblock


\bibitem[\protect\citeauthoryear{Jafri, Hassan, Hemani, and Mutlu}{Jafri
  et~al\mbox{.}}{2019}]%
        {jafri2019refresh}
\bibfield{author}{\bibinfo{person}{S.~M. Jafri}, {et~al\mbox{.}}}
  \bibinfo{year}{2019}\natexlab{}.
\newblock \showarticletitle{{Refresh Triggered Computation: Improving the
  Energy Efficiency of Convolutional Neural Network Accelerators}}. In
  \bibinfo{booktitle}{\emph{arXiv preprint arXiv:1910.06672}}.
\newblock


\bibitem[\protect\citeauthoryear{Jafri, Hemani, Paul, and Abbas}{Jafri
  et~al\mbox{.}}{2017}]%
        {mocha}
\bibfield{author}{\bibinfo{person}{S.~M. A.~H. Jafri}, {et~al\mbox{.}}}
  \bibinfo{year}{2017}\natexlab{}.
\newblock \showarticletitle{{MOCHA: Morphable Locality and Compression Aware
  Architecture for Convolutional Neural Networks}}. In
  \bibinfo{booktitle}{\emph{IPDPS}}.
\newblock


\bibitem[\protect\citeauthoryear{Jafri, Ozbak, Hemani, Farahini, Paul,
  et~al\mbox{.}}{Jafri et~al\mbox{.}}{2013}]%
        {asadisqed2012}
\bibfield{author}{\bibinfo{person}{S.~M. A.~H. Jafri}, {et~al\mbox{.}}}
  \bibinfo{year}{2013}\natexlab{}.
\newblock \showarticletitle{{Energy-Aware {CGRA}s using Dynamically
  Reconfigurable Isolation Cells}}. In \bibinfo{booktitle}{\emph{ISQED}}.
\newblock


\bibitem[\protect\citeauthoryear{{JEDEC}}{{JEDEC}}{2007}]%
        {standard2007ddr3}
\bibfield{author}{\bibinfo{person}{{JEDEC}}.} \bibinfo{year}{2007}\natexlab{}.
\newblock \showarticletitle{{DDR3 SDRAM Standard}}.
\newblock \bibinfo{journal}{\emph{JESD79-3}} (\bibinfo{year}{2007}).
\newblock


\bibitem[\protect\citeauthoryear{{JEDEC}}{{JEDEC}}{2014}]%
        {jedec-lpddr4}
\bibfield{author}{\bibinfo{person}{{JEDEC}}.} \bibinfo{year}{2014}\natexlab{}.
\newblock \bibinfo{title}{{Low Power Double Data Rate 4 (LPDDR4)}}.
\newblock \bibinfo{howpublished}{Standard No. JESD209-4}.
\newblock


\bibitem[\protect\citeauthoryear{Jeong, Yoon, Sunwoo, Sullivan, Lee,
  et~al\mbox{.}}{Jeong et~al\mbox{.}}{2012}]%
        {jeong2012balancing}
\bibfield{author}{\bibinfo{person}{M.~K. Jeong}, {et~al\mbox{.}}}
  \bibinfo{year}{2012}\natexlab{}.
\newblock \showarticletitle{{Balancing DRAM Locality and Parallelism in Shared
  Memory CMP Systems}}. In \bibinfo{booktitle}{\emph{HPCA}}.
\newblock


\bibitem[\protect\citeauthoryear{Jung, Zulian, Mathew, Herrmann, Brugger,
  et~al\mbox{.}}{Jung et~al\mbox{.}}{2015}]%
        {Jung-memsys2015}
\bibfield{author}{\bibinfo{person}{M.~Jung}, {et~al\mbox{.}}}
  \bibinfo{year}{2015}\natexlab{}.
\newblock \showarticletitle{{Omitting Refresh: A Case Study for Commodity and
  Wide I/O DRAMs}}. In \bibinfo{booktitle}{\emph{MEMSYS}}.
\newblock


\bibitem[\protect\citeauthoryear{Kang, Yu, Park, Zheng, Halbert,
  et~al\mbox{.}}{Kang et~al\mbox{.}}{2014}]%
        {kang-memcon2014}
\bibfield{author}{\bibinfo{person}{U.~Kang}, {et~al\mbox{.}}}
  \bibinfo{year}{2014}\natexlab{}.
\newblock \showarticletitle{{Co-Architecting Controllers and {DRAM} to Enhance
  {DRAM} Process Scaling}}. In \bibinfo{booktitle}{\emph{{The Memory Forum}}}.
\newblock


\bibitem[\protect\citeauthoryear{Keeth, Baker, Johnson, and Lin}{Keeth
  et~al\mbox{.}}{2007}]%
        {keeth2007dram}
\bibfield{author}{\bibinfo{person}{B.~Keeth}, {et~al\mbox{.}}}
  \bibinfo{year}{2007}\natexlab{}.
\newblock \bibinfo{booktitle}{\emph{{DRAM Circuit Design: Fundamental and
  High-Speed Topics}}}.
\newblock \bibinfo{publisher}{{John Wiley \& Sons}}.
\newblock


\bibitem[\protect\citeauthoryear{Khan, Lee, Kim, Alameldeen, Wilkerson,
  et~al\mbox{.}}{Khan et~al\mbox{.}}{2014}]%
        {khan-sigmetrics2014}
\bibfield{author}{\bibinfo{person}{S.~Khan}, {et~al\mbox{.}}}
  \bibinfo{year}{2014}\natexlab{}.
\newblock \showarticletitle{{The Efficacy of Error Mitigation Techniques for
  DRAM Retention Failures: A Comparative Experimental Study}}. In
  \bibinfo{booktitle}{\emph{SIGMETRICS}}.
\newblock


\bibitem[\protect\citeauthoryear{Khan, Lee, and Mutlu}{Khan
  et~al\mbox{.}}{2016a}]%
        {khan2016parbor}
\bibfield{author}{\bibinfo{person}{S.~Khan}, {et~al\mbox{.}}}
  \bibinfo{year}{2016}\natexlab{a}.
\newblock \showarticletitle{{{PARBOR}: An Efficient System-Level Technique to
  Detect Data-Dependent Failures in {DRAM}}}. In
  \bibinfo{booktitle}{\emph{DSN}}.
\newblock


\bibitem[\protect\citeauthoryear{Khan, Wilkerson, Lee, Alameldeen, and
  Mutlu}{Khan et~al\mbox{.}}{2016b}]%
        {khan2016case}
\bibfield{author}{\bibinfo{person}{S.~Khan}, {et~al\mbox{.}}}
  \bibinfo{year}{2016}\natexlab{b}.
\newblock \showarticletitle{{A Case for Memory Content-Based Detection and
  Mitigation of Data-Dependent Failures in {DRAM}}}. In
  \bibinfo{booktitle}{\emph{CAL}}.
\newblock


\bibitem[\protect\citeauthoryear{Khan, Wilkerson, Wang, Alameldeen, Lee,
  et~al\mbox{.}}{Khan et~al\mbox{.}}{2017}]%
        {khan2017detecting}
\bibfield{author}{\bibinfo{person}{S.~Khan}, {et~al\mbox{.}}}
  \bibinfo{year}{2017}\natexlab{}.
\newblock \showarticletitle{{Detecting and Mitigating Data-Dependent DRAM
  Failures by Exploiting Current Memory Content}}. In
  \bibinfo{booktitle}{\emph{MICRO}}.
\newblock


\bibitem[\protect\citeauthoryear{Kim and Kwon}{Kim and Kwon}{2020}]%
        {kim2020smart}
\bibfield{author}{\bibinfo{person}{D.-Y. Kim} {et~al\mbox{.}}}
  \bibinfo{year}{2020}\natexlab{}.
\newblock \showarticletitle{{Smart Adaptive Refresh for Optimum Refresh
  Interval Tracking using in-DRAM ECC}}. In \bibinfo{booktitle}{\emph{MWSCAS}}.
\newblock


\bibitem[\protect\citeauthoryear{Kim, Patel, Hassan, and Mutlu}{Kim
  et~al\mbox{.}}{2018a}]%
        {kim2018solar}
\bibfield{author}{\bibinfo{person}{J.~S. Kim}, {et~al\mbox{.}}}
  \bibinfo{year}{2018}\natexlab{a}.
\newblock \showarticletitle{{Solar-DRAM: Reducing DRAM Access Latency by
  Exploiting the Variation in Local Bitlines}}. In
  \bibinfo{booktitle}{\emph{ICCD}}.
\newblock


\bibitem[\protect\citeauthoryear{Kim, Patel, Hassan, and Mutlu}{Kim
  et~al\mbox{.}}{2018b}]%
        {kim2018dram}
\bibfield{author}{\bibinfo{person}{J.~S. Kim}, {et~al\mbox{.}}}
  \bibinfo{year}{2018}\natexlab{b}.
\newblock \showarticletitle{{The DRAM Latency PUF: Quickly Evaluating Physical
  Unclonable Functions by Exploiting the Latency-Reliability Tradeoff in Modern
  Commodity DRAM Devices}}. In \bibinfo{booktitle}{\emph{HPCA}}.
\newblock


\bibitem[\protect\citeauthoryear{Kim, Patel, Hassan, Orosa, and Mutlu}{Kim
  et~al\mbox{.}}{2019}]%
        {kim2019d}
\bibfield{author}{\bibinfo{person}{J.~S. Kim}, {et~al\mbox{.}}}
  \bibinfo{year}{2019}\natexlab{}.
\newblock \showarticletitle{{D-RaNGe: Using Commodity DRAM Devices to Generate
  True Random Numbers with Low Latency and High Throughput}}. In
  \bibinfo{booktitle}{\emph{HPCA}}.
\newblock


\bibitem[\protect\citeauthoryear{Kim, Patel, Ya{\u{g}}l{\i}k{\c{c}}{\i},
  Hassan, Azizi, et~al\mbox{.}}{Kim et~al\mbox{.}}{2020b}]%
        {kim2020revisiting}
\bibfield{author}{\bibinfo{person}{J.~S. Kim}, {et~al\mbox{.}}}
  \bibinfo{year}{2020}\natexlab{b}.
\newblock \showarticletitle{{Revisiting RowHammer: An Experimental Analysis of
  Modern Devices and Mitigation Techniques}}. In
  \bibinfo{booktitle}{\emph{ISCA}}.
\newblock


\bibitem[\protect\citeauthoryear{Kim, Kwak, Kim, Baek, and Huh}{Kim
  et~al\mbox{.}}{2020a}]%
        {kim2020charge}
\bibfield{author}{\bibinfo{person}{S.~Kim}, {et~al\mbox{.}}}
  \bibinfo{year}{2020}\natexlab{a}.
\newblock \showarticletitle{{Charge-Aware DRAM Refresh Reduction with Value
  Transformation}}. In \bibinfo{booktitle}{\emph{HPCA}}.
\newblock


\bibitem[\protect\citeauthoryear{Kim, Daly, Kim, Fallin, Lee,
  et~al\mbox{.}}{Kim et~al\mbox{.}}{2014}]%
        {kim2014flipping}
\bibfield{author}{\bibinfo{person}{Y.~Kim}, {et~al\mbox{.}}}
  \bibinfo{year}{2014}\natexlab{}.
\newblock \showarticletitle{{Flipping Bits in Memory without Accessing Them: An
  Experimental Study of DRAM Disturbance Errors}}. In
  \bibinfo{booktitle}{\emph{ISCA}}.
\newblock


\bibitem[\protect\citeauthoryear{Kim, Seshadri, Lee, Liu, and Mutlu}{Kim
  et~al\mbox{.}}{2012}]%
        {kim-isca2012}
\bibfield{author}{\bibinfo{person}{Y.~Kim}, {et~al\mbox{.}}}
  \bibinfo{year}{2012}\natexlab{}.
\newblock \showarticletitle{{A Case for Exploiting Subarray-Level Parallelism
  (SALP) in DRAM}}. In \bibinfo{booktitle}{\emph{ISCA}}.
\newblock


\bibitem[\protect\citeauthoryear{Kirby and Sirovich}{Kirby and
  Sirovich}{1990}]%
        {Kirby-TPAM1990}
\bibfield{author}{\bibinfo{person}{M.~Kirby} {et~al\mbox{.}}}
  \bibinfo{year}{1990}\natexlab{}.
\newblock \showarticletitle{{Application of the Karhunen-Loeve Procedure for
  the Characterization of Human Faces}}.
\newblock \bibinfo{journal}{\emph{IEEE TPAMI}}.
\newblock


\bibitem[\protect\citeauthoryear{Koppula, Orosa, Kanellopoulos, Shahroodi,
  Azizi, et~al\mbox{.}}{Koppula et~al\mbox{.}}{2019}]%
        {koppula2019eden}
\bibfield{author}{\bibinfo{person}{S.~Koppula}, {et~al\mbox{.}}}
  \bibinfo{year}{2019}\natexlab{}.
\newblock \showarticletitle{{EDEN: Energy-Efficient, High-Performance Neural
  Network Inference Using Approximate DRAM}}. In
  \bibinfo{booktitle}{\emph{MICRO}}.
\newblock


\bibitem[\protect\citeauthoryear{Krizhevsky, Sutskever, and Hinton}{Krizhevsky
  et~al\mbox{.}}{2012}]%
        {hinton2012imagenet}
\bibfield{author}{\bibinfo{person}{A.~Krizhevsky}, {et~al\mbox{.}}}
  \bibinfo{year}{2012}\natexlab{}.
\newblock \showarticletitle{{Imagenet Classification with Deep Convolutional
  Neural Networks}}. In \bibinfo{booktitle}{\emph{NIPS}}.
\newblock


\bibitem[\protect\citeauthoryear{Lavin and Gray}{Lavin and Gray}{2016}]%
        {lavin2016fast}
\bibfield{author}{\bibinfo{person}{A.~Lavin} {et~al\mbox{.}}}
  \bibinfo{year}{2016}\natexlab{}.
\newblock \showarticletitle{{Fast Algorithms for Convolutional Neural
  Networks}}. In \bibinfo{booktitle}{\emph{CoRR}}.
\newblock


\bibitem[\protect\citeauthoryear{Lecun, Bottou, Bengio, and Haffner}{Lecun
  et~al\mbox{.}}{1998}]%
        {lecun1998gradient}
\bibfield{author}{\bibinfo{person}{Y.~Lecun}, {et~al\mbox{.}}}
  \bibinfo{year}{1998}\natexlab{}.
\newblock \showarticletitle{{Gradient-based Learning Applied to Document
  Recognition}}.
\newblock \bibinfo{journal}{\emph{Proc. of the IEEE}}.
\newblock


\bibitem[\protect\citeauthoryear{Lee, Ghose, Pekhimenko, Khan, and Mutlu}{Lee
  et~al\mbox{.}}{2016}]%
        {lee2015simultaneous}
\bibfield{author}{\bibinfo{person}{D.~Lee}, {et~al\mbox{.}}}
  \bibinfo{year}{2016}\natexlab{}.
\newblock \showarticletitle{{Simultaneous Multi-Layer Access: Improving
  3D-Stacked Memory Bandwidth at Low Cost}}. In
  \bibinfo{booktitle}{\emph{{TACO}}}.
\newblock


\bibitem[\protect\citeauthoryear{Lee, Khan, Subramanian, Ghose,
  Ausavarungnirun, et~al\mbox{.}}{Lee et~al\mbox{.}}{2017}]%
        {lee2017design}
\bibfield{author}{\bibinfo{person}{D.~Lee}, {et~al\mbox{.}}}
  \bibinfo{year}{2017}\natexlab{}.
\newblock \showarticletitle{{Design-Induced Latency Variation in Modern DRAM
  Chips: Characterization, Analysis, and Latency Reduction Mechanisms}}. In
  \bibinfo{booktitle}{\emph{SIGMETRICS}}.
\newblock


\bibitem[\protect\citeauthoryear{Lee, Kim, Pekhimenko, Khan, Seshadri,
  et~al\mbox{.}}{Lee et~al\mbox{.}}{2015a}]%
        {lee-hpca2015}
\bibfield{author}{\bibinfo{person}{D.~Lee}, {et~al\mbox{.}}}
  \bibinfo{year}{2015}\natexlab{a}.
\newblock \showarticletitle{{Adaptive-Latency DRAM: Optimizing DRAM Timing for
  the Common-Case}}. In \bibinfo{booktitle}{\emph{HPCA}}.
\newblock


\bibitem[\protect\citeauthoryear{Lee, Kim, Seshadri, Liu, Subramanian,
  et~al\mbox{.}}{Lee et~al\mbox{.}}{2013}]%
        {lee-hpca2013}
\bibfield{author}{\bibinfo{person}{D.~Lee}, {et~al\mbox{.}}}
  \bibinfo{year}{2013}\natexlab{}.
\newblock \showarticletitle{{Tiered-Latency DRAM: A Low Latency and Low Cost
  DRAM Architecture}}. In \bibinfo{booktitle}{\emph{HPCA}}.
\newblock


\bibitem[\protect\citeauthoryear{Lee, Subramanian, Ausavarungnirun, Choi, and
  Mutlu}{Lee et~al\mbox{.}}{2015b}]%
        {lee2015decoupled}
\bibfield{author}{\bibinfo{person}{D.~Lee}, {et~al\mbox{.}}}
  \bibinfo{year}{2015}\natexlab{b}.
\newblock \showarticletitle{{Decoupled Direct Memory Access: Isolating CPU and
  IO Traffic by Leveraging a Dual-Data-Port DRAM}}. In
  \bibinfo{booktitle}{\emph{PACT}}.
\newblock


\bibitem[\protect\citeauthoryear{Leupers and Marwedel}{Leupers and
  Marwedel}{1996}]%
        {leupers1996algorithms}
\bibfield{author}{\bibinfo{person}{R.~Leupers} {et~al\mbox{.}}}
  \bibinfo{year}{1996}\natexlab{}.
\newblock \showarticletitle{{Algorithms for Address Assignment in DSP Code
  Generation}}. In \bibinfo{booktitle}{\emph{ICCAD}}.
\newblock


\bibitem[\protect\citeauthoryear{Liu}{Liu}{2008}]%
        {liu2008embedded}
\bibfield{author}{\bibinfo{person}{D.~Liu}.} \bibinfo{year}{2008}\natexlab{}.
\newblock \bibinfo{booktitle}{\emph{{Embedded DSP Processor Design: Application
  Specific Instruction Set Processors}}}.
\newblock \bibinfo{publisher}{Elsevier}.
\newblock


\bibitem[\protect\citeauthoryear{Liu et~al\mbox{.}}{Liu et~al\mbox{.}}{2012b}]%
        {liu2012raidr}
\bibfield{author}{\bibinfo{person}{J.~Liu} {et~al\mbox{.}}}
  \bibinfo{year}{2012}\natexlab{b}.
\newblock \showarticletitle{{{RAIDR: Retention-Aware} Intelligent {DRAM}
  Refresh}}. In \bibinfo{booktitle}{\emph{ISCA}}.
\newblock


\bibitem[\protect\citeauthoryear{Liu et~al\mbox{.}}{Liu et~al\mbox{.}}{2013}]%
        {liu2013experimental}
\bibfield{author}{\bibinfo{person}{J.~Liu} {et~al\mbox{.}}}
  \bibinfo{year}{2013}\natexlab{}.
\newblock \showarticletitle{{An Experimental Study of Data Retention Behavior
  in Modern {DRAM} Devices: Implications for Retention Time Profiling
  Mechanisms}}. In \bibinfo{booktitle}{\emph{ISCA}}.
\newblock


\bibitem[\protect\citeauthoryear{Liu, Cui, Xing, Bao, Chen, et~al\mbox{.}}{Liu
  et~al\mbox{.}}{2012a}]%
        {liu2012software}
\bibfield{author}{\bibinfo{person}{L.~Liu}, {et~al\mbox{.}}}
  \bibinfo{year}{2012}\natexlab{a}.
\newblock \showarticletitle{{A Software Memory Partition Approach for
  Eliminating Bank-level Interference in Multicore Systems}}. In
  \bibinfo{booktitle}{\emph{PACT}}.
\newblock


\bibitem[\protect\citeauthoryear{Liu, Pattabiraman, Moscibroda, and Zorn}{Liu
  et~al\mbox{.}}{2011}]%
        {Liu:2011}
\bibfield{author}{\bibinfo{person}{S.~Liu}, {et~al\mbox{.}}}
  \bibinfo{year}{2011}\natexlab{}.
\newblock \showarticletitle{{Flikker: Saving DRAM Refresh-power Through
  Critical Data Partitioning}}. In \bibinfo{booktitle}{\emph{ASPLOS}}.
\newblock


\bibitem[\protect\citeauthoryear{Luo, Shahroodi, Hassan, Patel, Yaglikci,
  et~al\mbox{.}}{Luo et~al\mbox{.}}{2020}]%
        {luo2020clr}
\bibfield{author}{\bibinfo{person}{H.~Luo}, {et~al\mbox{.}}}
  \bibinfo{year}{2020}\natexlab{}.
\newblock \showarticletitle{{CLR-DRAM: A Low-Cost DRAM Architecture Enabling
  Dynamic Capacity-Latency Trade-Off}}. In \bibinfo{booktitle}{\emph{ISCA}}.
\newblock


\bibitem[\protect\citeauthoryear{Luo, Govindan, Sharma, Santaniello, Meza,
  et~al\mbox{.}}{Luo et~al\mbox{.}}{2014}]%
        {luo2014characterizing}
\bibfield{author}{\bibinfo{person}{Y.~Luo}, {et~al\mbox{.}}}
  \bibinfo{year}{2014}\natexlab{}.
\newblock \showarticletitle{{Characterizing Application Memory Error
  Vulnerability to Optimize Datacenter Cost via Heterogeneous-Reliability
  Memory}}. In \bibinfo{booktitle}{\emph{DSN}}.
\newblock


\bibitem[\protect\citeauthoryear{Malladi, Shaeffer, Gopalakrishnan, Lo, Lee,
  et~al\mbox{.}}{Malladi et~al\mbox{.}}{2012}]%
        {Mall-micro2012}
\bibfield{author}{\bibinfo{person}{K.~T. Malladi}, {et~al\mbox{.}}}
  \bibinfo{year}{2012}\natexlab{}.
\newblock \showarticletitle{{Rethinking DRAM Power Modes for Energy
  Proportionality}}. In \bibinfo{booktitle}{\emph{MICRO}}.
\newblock


\bibitem[\protect\citeauthoryear{Mathew and Davis}{Mathew and Davis}{2004}]%
        {mathew2004loop}
\bibfield{author}{\bibinfo{person}{B.~Mathew} {et~al\mbox{.}}}
  \bibinfo{year}{2004}\natexlab{}.
\newblock \showarticletitle{{A Loop Accelerator for Low Power Embedded VLIW
  Processors}}. In \bibinfo{booktitle}{\emph{CODES}}.
\newblock


\bibitem[\protect\citeauthoryear{Micron}{Micron}{2014}]%
        {micron_lpddr3}
\bibfield{author}{\bibinfo{person}{Micron}.} \bibinfo{year}{2014}\natexlab{}.
\newblock \bibinfo{title}{{Mobile LPDDR3 SDRAM}}.
\newblock
\newblock
\urldef\tempurl%
\url{https://www.micron.com/-/media/client/global/documents/products/data-sheet/dram/mobile-dram/low-power-dram/lpddr3/253b_12-5x12-5_2ch_8-16gb_2c0f_mobile_lpddr3.pdf?rev=1b66d5710434460eb13dc3be8faa6d77}
\showURL{%
\tempurl}


\bibitem[\protect\citeauthoryear{Mukundan, Hunter, Kim, Stuecheli, and
  Mart{\'\i}nez}{Mukundan et~al\mbox{.}}{2013}]%
        {mukundan2013understanding}
\bibfield{author}{\bibinfo{person}{J.~Mukundan}, {et~al\mbox{.}}}
  \bibinfo{year}{2013}\natexlab{}.
\newblock \showarticletitle{{Understanding and Mitigating Refresh Overheads in
  High-Density DDR4 DRAM Systems}}. In \bibinfo{booktitle}{\emph{ISCA}}.
\newblock


\bibitem[\protect\citeauthoryear{Muralidhara, Subramanian, Mutlu, Kandemir, and
  Moscibroda}{Muralidhara et~al\mbox{.}}{2011}]%
        {muralidhara2011reducing}
\bibfield{author}{\bibinfo{person}{S.~P. Muralidhara}, {et~al\mbox{.}}}
  \bibinfo{year}{2011}\natexlab{}.
\newblock \showarticletitle{{Reducing Memory Interference in Multicore Systems
  via Application-Aware Memory Channel Partitioning}}. In
  \bibinfo{booktitle}{\emph{MICRO}}.
\newblock


\bibitem[\protect\citeauthoryear{Mutlu}{Mutlu}{2013}]%
        {mutlu2013memory}
\bibfield{author}{\bibinfo{person}{O.~Mutlu}.} \bibinfo{year}{2013}\natexlab{}.
\newblock \showarticletitle{{Memory Scaling: {A} Systems Architecture
  Perspective}}.
\newblock \bibinfo{journal}{\emph{IMW}}.
\newblock


\bibitem[\protect\citeauthoryear{Mutlu}{Mutlu}{2017}]%
        {mutlu2017rowhammer}
\bibfield{author}{\bibinfo{person}{O.~Mutlu}.} \bibinfo{year}{2017}\natexlab{}.
\newblock \showarticletitle{{The RowHammer Problem and Other Issues we may Face
  as Memory Becomes Denser}}. In \bibinfo{booktitle}{\emph{DATE}}.
\newblock


\bibitem[\protect\citeauthoryear{Mutlu and Kim}{Mutlu and Kim}{2019}]%
        {mutlu2019rowhammer}
\bibfield{author}{\bibinfo{person}{O.~Mutlu} {et~al\mbox{.}}}
  \bibinfo{year}{2019}\natexlab{}.
\newblock \showarticletitle{{RowHammer: A Retrospective}}. In
  \bibinfo{booktitle}{\emph{TCAD}}.
\newblock


\bibitem[\protect\citeauthoryear{Mutlu and Moscibroda}{Mutlu and
  Moscibroda}{2008}]%
        {mutlu2008parallelism}
\bibfield{author}{\bibinfo{person}{O.~Mutlu} {et~al\mbox{.}}}
  \bibinfo{year}{2008}\natexlab{}.
\newblock \showarticletitle{{Parallelism-Aware Batch Scheduling: Enhancing Both
  Performance and Fairness of Shared DRAM Systems}}. In
  \bibinfo{booktitle}{\emph{ISCA}}.
\newblock


\bibitem[\protect\citeauthoryear{Mutlu and Subramanian}{Mutlu and
  Subramanian}{2015}]%
        {mutlu2015research}
\bibfield{author}{\bibinfo{person}{O.~Mutlu} {et~al\mbox{.}}}
  \bibinfo{year}{2015}\natexlab{}.
\newblock \showarticletitle{{Research Problems and Opportunities in Memory
  Systems}}.
\newblock \bibinfo{journal}{\emph{SUPERFRI}}.
\newblock


\bibitem[\protect\citeauthoryear{Nair, Chou, and Qureshi}{Nair
  et~al\mbox{.}}{2013a}]%
        {nair2013case}
\bibfield{author}{\bibinfo{person}{P.~Nair}, {et~al\mbox{.}}}
  \bibinfo{year}{2013}\natexlab{a}.
\newblock \showarticletitle{{A Case for Refresh Pausing in DRAM Memory
  Systems}}. In \bibinfo{booktitle}{\emph{HPCA}}.
\newblock


\bibitem[\protect\citeauthoryear{Nair, Chou, and Qureshi}{Nair
  et~al\mbox{.}}{2014}]%
        {nair2014refresh}
\bibfield{author}{\bibinfo{person}{P.~J. Nair}, {et~al\mbox{.}}}
  \bibinfo{year}{2014}\natexlab{}.
\newblock \showarticletitle{{Refresh Pausing in DRAM Memory Systems}}. In
  \bibinfo{booktitle}{\emph{TACO}}.
\newblock


\bibitem[\protect\citeauthoryear{Nair, Kim, and Qureshi}{Nair
  et~al\mbox{.}}{2013b}]%
        {nair2013archshield}
\bibfield{author}{\bibinfo{person}{P.~J. Nair}, {et~al\mbox{.}}}
  \bibinfo{year}{2013}\natexlab{b}.
\newblock \showarticletitle{{ArchShield: Architectural Framework for Assisting
  DRAM Scaling by Tolerating High Error Rates}}. In
  \bibinfo{booktitle}{\emph{ISCA}}.
\newblock


\bibitem[\protect\citeauthoryear{Nair, Sridharan, and Qureshi}{Nair
  et~al\mbox{.}}{2016}]%
        {nair2016xed}
\bibfield{author}{\bibinfo{person}{P.~J. Nair}, {et~al\mbox{.}}}
  \bibinfo{year}{2016}\natexlab{}.
\newblock \showarticletitle{{XED: Exposing On-Die Error Detection Information
  for Strong Memory Reliability}}. In \bibinfo{booktitle}{\emph{ISCA}}.
\newblock


\bibitem[\protect\citeauthoryear{Patel, Kim, Hassan, and Mutlu}{Patel
  et~al\mbox{.}}{2019}]%
        {patel2019understanding}
\bibfield{author}{\bibinfo{person}{M.~Patel}, {et~al\mbox{.}}}
  \bibinfo{year}{2019}\natexlab{}.
\newblock \showarticletitle{{Understanding and Modeling On-Die Error Correction
  in Modern DRAM: An Experimental Study Using Real Devices}}. In
  \bibinfo{booktitle}{\emph{DSN}}.
\newblock


\bibitem[\protect\citeauthoryear{Patel, Kim, and Mutlu}{Patel
  et~al\mbox{.}}{2017}]%
        {reaper}
\bibfield{author}{\bibinfo{person}{M.~Patel}, {et~al\mbox{.}}}
  \bibinfo{year}{2017}\natexlab{}.
\newblock \showarticletitle{{The Reach Profiler (REAPER): Enabling the
  Mitigation of DRAM Retention Failures via Profiling at Aggressive
  Conditions}}. In \bibinfo{booktitle}{\emph{ISCA}}.
\newblock


\bibitem[\protect\citeauthoryear{Patel, Kim, Shahroodi, Hassan, and
  Mutlu}{Patel et~al\mbox{.}}{2020}]%
        {patel2020bit}
\bibfield{author}{\bibinfo{person}{M.~Patel}, {et~al\mbox{.}}}
  \bibinfo{year}{2020}\natexlab{}.
\newblock \showarticletitle{{Bit-Exact ECC Recovery (BEER): Determining DRAM
  On-Die ECC Functions by Exploiting DRAM Data Retention Characteristics}}.
\newblock \bibinfo{journal}{\emph{MICRO}}.
\newblock


\bibitem[\protect\citeauthoryear{Pourshirazi and Zhu}{Pourshirazi and
  Zhu}{2016}]%
        {pourshirazi2016refree}
\bibfield{author}{\bibinfo{person}{B.~Pourshirazi} {et~al\mbox{.}}}
  \bibinfo{year}{2016}\natexlab{}.
\newblock \showarticletitle{{Refree: A Refresh-free Hybrid DRAM/PCM Main Memory
  System}}. In \bibinfo{booktitle}{\emph{IPDPS}}.
\newblock


\bibitem[\protect\citeauthoryear{Qadeer, Hameed, Shacham, Venkatesan,
  Kozyrakis, et~al\mbox{.}}{Qadeer et~al\mbox{.}}{2013}]%
        {Qadeer-ISCA2013}
\bibfield{author}{\bibinfo{person}{W.~Qadeer}, {et~al\mbox{.}}}
  \bibinfo{year}{2013}\natexlab{}.
\newblock \showarticletitle{{Convolution Engine: Balancing Efficiency \&
  Flexibility in Specialized Computing}}. In \bibinfo{booktitle}{\emph{ISCA}}.
\newblock


\bibitem[\protect\citeauthoryear{Qureshi, Kim, Khan, Nair, and Mutlu}{Qureshi
  et~al\mbox{.}}{2015}]%
        {qureshi-dsn2015}
\bibfield{author}{\bibinfo{person}{M.~Qureshi}, {et~al\mbox{.}}}
  \bibinfo{year}{2015}\natexlab{}.
\newblock \showarticletitle{{AVATAR: A Variable-Retention-Time (VRT) Aware
  Refresh for DRAM Systems}}. In \bibinfo{booktitle}{\emph{DSN}}.
\newblock


\bibitem[\protect\citeauthoryear{Rumelhart, Hinton, and Williams}{Rumelhart
  et~al\mbox{.}}{1985}]%
        {rumelhart1985learning}
\bibfield{author}{\bibinfo{person}{D.~E. Rumelhart}, {et~al\mbox{.}}}
  \bibinfo{year}{1985}\natexlab{}.
\newblock \bibinfo{booktitle}{\emph{{Learning Internal Representations by Error
  Propagation}}}.
\newblock \bibinfo{type}{{T}echnical {R}eport}.
\newblock


\bibitem[\protect\citeauthoryear{Seshadri, Kim, Fallin, Lee, Ausavarungnirun,
  et~al\mbox{.}}{Seshadri et~al\mbox{.}}{2013}]%
        {seshadri2013rowclone}
\bibfield{author}{\bibinfo{person}{V.~Seshadri}, {et~al\mbox{.}}}
  \bibinfo{year}{2013}\natexlab{}.
\newblock \showarticletitle{{RowClone: Fast and Energy-Efficient In-DRAM Bulk
  Data Copy and Initialization}}. In \bibinfo{booktitle}{\emph{MICRO}}.
\newblock


\bibitem[\protect\citeauthoryear{Seshadri, Lee, Mullins, Hassan, Boroumand,
  et~al\mbox{.}}{Seshadri et~al\mbox{.}}{2017}]%
        {seshadri2017ambit}
\bibfield{author}{\bibinfo{person}{V.~Seshadri}, {et~al\mbox{.}}}
  \bibinfo{year}{2017}\natexlab{}.
\newblock \showarticletitle{{Ambit: In-Memory Accelerator for Bulk Bitwise
  Operations Using Commodity DRAM Technology}}. In
  \bibinfo{booktitle}{\emph{MICRO}}.
\newblock


\bibitem[\protect\citeauthoryear{Seshadri and Mutlu}{Seshadri and
  Mutlu}{2019}]%
        {seshadri2019dram}
\bibfield{author}{\bibinfo{person}{V.~Seshadri} {et~al\mbox{.}}}
  \bibinfo{year}{2019}\natexlab{}.
\newblock \showarticletitle{{In-DRAM Bulk Bitwise Execution Engine}}.
\newblock \bibinfo{journal}{\emph{arXiv preprint arXiv:1905.09822}}
  (\bibinfo{year}{2019}).
\newblock


\bibitem[\protect\citeauthoryear{Shi, Xu, Sun, Peemen, Li, et~al\mbox{.}}{Shi
  et~al\mbox{.}}{2015}]%
        {shi2015a}
\bibfield{author}{\bibinfo{person}{R.~Shi}, {et~al\mbox{.}}}
  \bibinfo{year}{2015}\natexlab{}.
\newblock \showarticletitle{{A Locality Aware Convolutional Neural Networks
  Accelerator}}. In \bibinfo{booktitle}{\emph{DSD}}.
\newblock


\bibitem[\protect\citeauthoryear{Smith, Waterman, et~al\mbox{.}}{Smith
  et~al\mbox{.}}{1981}]%
        {smith1981identification}
\bibfield{author}{\bibinfo{person}{T.~F. Smith}, {et~al\mbox{.}}}
  \bibinfo{year}{1981}\natexlab{}.
\newblock \showarticletitle{{Identification of Common Molecular Subsequences}}.
\newblock \bibinfo{journal}{\emph{Journal of molecular biology}}.
\newblock


\bibitem[\protect\citeauthoryear{Song, Xingyu, Huizi, Jing, Ardavan,
  et~al\mbox{.}}{Song et~al\mbox{.}}{2016}]%
        {dally2016eie}
\bibfield{author}{\bibinfo{person}{H.~Song}, {et~al\mbox{.}}}
  \bibinfo{year}{2016}\natexlab{}.
\newblock \showarticletitle{{EIE: Efficient Inference Engine on Compressed Deep
  Neural Network}}. In \bibinfo{booktitle}{\emph{ISCA}}.
\newblock


\bibitem[\protect\citeauthoryear{Stuecheli, Kaseridis, Hunter, and
  John}{Stuecheli et~al\mbox{.}}{2010}]%
        {stue-micro2010}
\bibfield{author}{\bibinfo{person}{J.~Stuecheli}, {et~al\mbox{.}}}
  \bibinfo{year}{2010}\natexlab{}.
\newblock \showarticletitle{{Elastic Refresh: Techniques to Mitigate Refresh
  Penalties in High Density Memory}}. In \bibinfo{booktitle}{\emph{MICRO}}.
\newblock


\bibitem[\protect\citeauthoryear{Szegedy, Liu, Jia, Sermanet, Reed,
  et~al\mbox{.}}{Szegedy et~al\mbox{.}}{2015}]%
        {szegedy2015going}
\bibfield{author}{\bibinfo{person}{C.~Szegedy}, {et~al\mbox{.}}}
  \bibinfo{year}{2015}\natexlab{}.
\newblock \showarticletitle{{Going Deeper with Convolutions}}. In
  \bibinfo{booktitle}{\emph{CVPR}}.
\newblock


\bibitem[\protect\citeauthoryear{Taniguchi, Jayapala, Raghavan, Catthoor,
  Sakanushi, et~al\mbox{.}}{Taniguchi et~al\mbox{.}}{2009a}]%
        {taniguchi2009systematic}
\bibfield{author}{\bibinfo{person}{I.~Taniguchi}, {et~al\mbox{.}}}
  \bibinfo{year}{2009}\natexlab{a}.
\newblock \showarticletitle{{Systematic Architecture Exploration Based on
  Optimistic Cycle Estimation for Low Energy Embedded Processors}}. In
  \bibinfo{booktitle}{\emph{ASP-DAC}}.
\newblock


\bibitem[\protect\citeauthoryear{Taniguchi, Raghavan, Jayapala, Catthoor,
  Takeuchi, et~al\mbox{.}}{Taniguchi et~al\mbox{.}}{2009b}]%
        {taniguchi2009reconfigurable}
\bibfield{author}{\bibinfo{person}{I.~Taniguchi}, {et~al\mbox{.}}}
  \bibinfo{year}{2009}\natexlab{b}.
\newblock \showarticletitle{{Reconfigurable AGU: An Address Generation Unit
  Based on Address Calculation Pattern for Low Energy and High Performance
  Embedded Processors}}.
\newblock \bibinfo{journal}{\emph{IEICE}}.
\newblock


\bibitem[\protect\citeauthoryear{Temam}{Temam}{2012}]%
        {Temam-ISCA2012}
\bibfield{author}{\bibinfo{person}{O.~Temam}.} \bibinfo{year}{2012}\natexlab{}.
\newblock \showarticletitle{{A Defect-tolerant Accelerator for Emerging
  High-performance Applications}}. In \bibinfo{booktitle}{\emph{ISCA}}.
\newblock


\bibitem[\protect\citeauthoryear{TN-46-15}{TN-46-15}{2007}]%
        {lowpowerddr}
\bibfield{author}{\bibinfo{person}{TN-46-15}.} \bibinfo{year}{2007}\natexlab{}.
\newblock \bibinfo{booktitle}{\emph{{Low-Power Versus Standard DDR SDRAM}}}.
\newblock \bibinfo{type}{{T}echnical {R}eport}. \bibinfo{institution}{Micron
  Technology, Inc.}
\newblock


\bibitem[\protect\citeauthoryear{Tu, Wu, Yin, Liu, and Wei}{Tu
  et~al\mbox{.}}{2018}]%
        {tu2018rana}
\bibfield{author}{\bibinfo{person}{F.~Tu}, {et~al\mbox{.}}}
  \bibinfo{year}{2018}\natexlab{}.
\newblock \showarticletitle{{RANA: Towards Efficient Neural Acceleration with
  Refresh-Optimized Embedded DRAM}}. In \bibinfo{booktitle}{\emph{ISCA}}.
\newblock


\bibitem[\protect\citeauthoryear{Udayanarayanan and Chakrabarti}{Udayanarayanan
  and Chakrabarti}{2001}]%
        {udayanarayanan2001address}
\bibfield{author}{\bibinfo{person}{S.~Udayanarayanan} {et~al\mbox{.}}}
  \bibinfo{year}{2001}\natexlab{}.
\newblock \showarticletitle{{Address Code Generation for Digital Signal
  Processors}}. In \bibinfo{booktitle}{\emph{DAC}}.
\newblock


\bibitem[\protect\citeauthoryear{Velilla}{Velilla}{2009}]%
        {velilla2009scratchpad}
\bibfield{author}{\bibinfo{person}{G.~T. Velilla}.}
  \bibinfo{year}{2009}\natexlab{}.
\newblock \emph{\bibinfo{title}{{Scratchpad-Oriented Address Generation for
  Low-Power Embedded VLIW Processors}}}.
\newblock \bibinfo{thesistype}{Ph.D. Dissertation}.
\newblock


\bibitem[\protect\citeauthoryear{Venkatesan et~al\mbox{.}}{Venkatesan
  et~al\mbox{.}}{2006}]%
        {venkatesan2006retention}
\bibfield{author}{\bibinfo{person}{R.~Venkatesan} {et~al\mbox{.}}}
  \bibinfo{year}{2006}\natexlab{}.
\newblock \showarticletitle{{Retention-Aware Placement in {DRAM (RAPID)}:
  {Software} Methods for Quasi-Non-Volatile {DRAM}}}. In
  \bibinfo{booktitle}{\emph{HPCA}}.
\newblock


\bibitem[\protect\citeauthoryear{Vogelsang}{Vogelsang}{2010}]%
        {Vogelsang-micro2010}
\bibfield{author}{\bibinfo{person}{T.~Vogelsang}.}
  \bibinfo{year}{2010}\natexlab{}.
\newblock \showarticletitle{{Understanding the Energy Consumption of Dynamic
  Random Access Memories}}. In \bibinfo{booktitle}{\emph{MICRO}}.
\newblock


\bibitem[\protect\citeauthoryear{Wang, Orosa, Peng, Guo, Ghose,
  et~al\mbox{.}}{Wang et~al\mbox{.}}{2020}]%
        {wang2020figaro}
\bibfield{author}{\bibinfo{person}{Y.~Wang}, {et~al\mbox{.}}}
  \bibinfo{year}{2020}\natexlab{}.
\newblock \showarticletitle{{FIGARO: Improving System Performance via
  Fine-Grained In-DRAM Data Relocation and Caching}}. In
  \bibinfo{booktitle}{\emph{MICRO}}.
\newblock


\bibitem[\protect\citeauthoryear{Wang, Tavakkol, Orosa, Ghose, Ghiasi,
  et~al\mbox{.}}{Wang et~al\mbox{.}}{2018}]%
        {wang2018reducing}
\bibfield{author}{\bibinfo{person}{Y.~Wang}, {et~al\mbox{.}}}
  \bibinfo{year}{2018}\natexlab{}.
\newblock \showarticletitle{{Reducing DRAM Latency via Charge-Level-Aware
  Look-Ahead Partial Restoration}}. In \bibinfo{booktitle}{\emph{MICRO}}.
\newblock


\bibitem[\protect\citeauthoryear{Xue, Shao, Chen, and Sha}{Xue
  et~al\mbox{.}}{2005}]%
        {xue2005optimizing}
\bibfield{author}{\bibinfo{person}{C.~Xue}, {et~al\mbox{.}}}
  \bibinfo{year}{2005}\natexlab{}.
\newblock \showarticletitle{{Optimizing DSP Scheduling via Address Assignment
  with Array and Loop Transformation}}. In \bibinfo{booktitle}{\emph{ICASSP}}.
\newblock


\bibitem[\protect\citeauthoryear{Yang, Stathis, Sharma, Paul, Hemani,
  et~al\mbox{.}}{Yang et~al\mbox{.}}{2018}]%
        {yang2018ribosom}
\bibfield{author}{\bibinfo{person}{Y.~Yang}, {et~al\mbox{.}}}
  \bibinfo{year}{2018}\natexlab{}.
\newblock \showarticletitle{{RiBoSOM: Rapid Bacterial Genome Identification
  Using Self-Organizing Map Implemented on the Synchoros SiLago Platform}}. In
  \bibinfo{booktitle}{\emph{SAMOS}}.
\newblock


\bibitem[\protect\citeauthoryear{Zhang, Chen, Xu, Sun, Wang,
  et~al\mbox{.}}{Zhang et~al\mbox{.}}{2014}]%
        {zhang2014half}
\bibfield{author}{\bibinfo{person}{T.~Zhang}, {et~al\mbox{.}}}
  \bibinfo{year}{2014}\natexlab{}.
\newblock \showarticletitle{{Half-DRAM: A High-Bandwidth and Low-Power DRAM
  Architecture from the Rethinking of Fine-Grained Activation}}. In
  \bibinfo{booktitle}{\emph{ISCA}}.
\newblock


\bibitem[\protect\citeauthoryear{Zulian, Weis, and Wehn}{Zulian
  et~al\mbox{.}}{2020}]%
        {zulian2020access}
\bibfield{author}{\bibinfo{person}{{\'E}.~F. Zulian}, {et~al\mbox{.}}}
  \bibinfo{year}{2020}\natexlab{}.
\newblock \showarticletitle{{Access-Aware Per-Bank DRAM Refresh for Reduced
  DRAM Refresh Overhead}}. In \bibinfo{booktitle}{\emph{ISCAS}}.
\newblock


\end{thebibliography}
%%%%%%%%%%%%%%%%%%%%%%%%%%%%%%%%%%%%

\end{document}